\documentclass[twocolumn,prl,superscriptaddress,floatfix,aps,10pt,showpacs]{revtex4-1}

\usepackage{graphicx}
\usepackage{amssymb}
\usepackage{amsmath}
\usepackage{multirow}
\usepackage{array,tabularx}
\usepackage{xcolor}

\newcommand{\vect}[1]{{\mathbf #1}}
\newcommand{\Frac}[2]{\displaystyle\frac{#1}{#2}}

\usepackage{float}
\usepackage{upgreek}
\usepackage{makeidx}
\DeclareGraphicsExtensions{.png,.jpg,.pdf,.mps,.gif,.bmp,.eps}
\graphicspath{{figures/}}
\usepackage{physics}

\renewcommand\zeta{\xi} 

\usepackage{lipsum}

\usepackage[colorlinks,citecolor=blue,urlcolor=blue]{hyperref}
\usepackage{cleveref}

\begin{document}
	
	\title{Unconventional Berezinskii-Kosterlitz-Thouless Transition in the Multicomponent Polariton System}
		
	\author{G. Dagvadorj} 
	\affiliation{Department of Physics, University of Warwick, Coventry,
		CV4 7AL, UK}
	\affiliation{Department of Physics and Astronomy, University College
			London, Gower Street, London, WC1E 6BT, UK}
	
	\author{P. Comaron}
	\email[Corresponding author: ]{p.comaron@ucl.ac.uk} 
	\affiliation{Department of Physics and Astronomy, University College
		London, Gower Street, London, WC1E 6BT, UK}
	\affiliation{Institute of Physics, Polish Academy of Sciences, Al. Lotnik\'ow 32/46, 02-668 Warsaw, Poland}
	
	\author{M.~H.~Szyma\'nska} 
	\email[Corresponding author: ]{m.szymanska@ucl.ac.uk}
	\affiliation{Department of Physics and Astronomy, University College
		London, Gower Street, London, WC1E 6BT, UK}

	\begin{abstract}
		We study a four-component  polariton system in the optical parametric oscillator regime consisting of exciton/photon and signal/idler modes
		across the Berezinskii-Kosterlitz-Thouless (BKT) transition.
		We show that all four components share the same BKT critical point, and algebraic decay of spatial coherence with the same critical exponent.
		However, while the collective excitations in different components are strongly locked, both close to and far from criticality, the spontaneous creation of topological defects in the vicinity of the phase transition is  found to be largely independent of the  {inter}-component mode locking, and instead strongly dependent on the density within a given mode. 
		 {This} peculiar characteristic allows us to reveal a novel state of matter, characterised by {configurations} of topological defects proliferating on top of a superfluid with algebraic decay of coherence, observation of which is demonstrated to be within  {reach of current experiments}.
	\end{abstract}
	
	
	\maketitle
	
	The study of nonequilibrium phase transitions in driven-dissipative quantum systems became of particular interest in recent years due to the unprecedented experimental progress  {in} realising quantum fluids of light.
	In two dimensions,  {the} onset of order is described by the celebrated Berezinskii-Kosterlitz-Thouless mechanism \cite{Berezinskii1971,Berezinskii1973}, where the quasi-condensate phase energy is minimised by  {the} pairing of topological defects with opposite charge.
	Such a  {phenomenon} has been investigated in a wide class of conservative systems, from $^4$He \cite{Bishop1978} to ultracold atoms \cite{Hadzibabic2006connecting}, and in different types of confinements~\cite{Hadzibabic2006}. 
	In the context of driven-dissipative polariton  {fluids}~\cite{kasprzak2006bose,carusotto2013quantum}, numerical studies of incoherently  {driven} {\cite{Comaron2021,Mei2021}}, as well as coherently driven systems in the OPO regime \cite{dagvadorj2015nonequilibrium}, have predicted a non-equilibrium-type BKT phase transition, recently confirmed in an experiment~\cite{caputo2018}.
	{Finally, the presence of further qualitative corrections to the standard BKT picture introduced by higher-order phase fluctuations treated within the Kardar–Parisi–Zhang framework has been proposed \cite{KPZ1986,altman2015twodimensional,fontaine2021,Ferrier2022}.}

	The polariton fluids are, however, intrinsically multicomponent. They consist of excitonic, photonic and, in the case of the OPO regime, the signal, idler and pump modes. Within the OPO picture, the polariton coherence  has been previously studied theoretically \cite{ciuti2001parametric,ciutiOPOtheory,carusotto2005spontaneous,wouters2007parametric,dunnett2016keldysh,dunnett2018properties} and experimentally~\cite{Krizhanovskii2006,Baumberg2010}.
	At a mean-field level the sum of the signal and idler phases is locked to the spatially-coherent pump phase~\cite{carusotto2013quantum}. However, including fluctuations {such a phase locking does not hold or is weak, especially in the vicinity of the critical point}~\cite{dagvadorj2015nonequilibrium}. 
	Despite that, the investigations to date were limited to  {the} signal mode of the photonic component~\cite{dagvadorj2015nonequilibrium} only, while 
	the interplay of critical features in different components are yet unrevealed.
	Therefore, questions regarding the multicomponent nature of the phase transition remains open:
	(i) How the spatial coherence differs in the different modes,  and how this affects the BKT phase transition?
	(ii) How topological defects in distinct components are correlated?

	In this work, we  investigate numerically the OPO thresholds
	showing that the nonequilibrium BKT transition occurs simultaneously (for the same pump strength) in all
	four components (photonic, excitonic, signal and idler) despite marked differences in the density of the different components, 
	Moreover, we demonstrate that coherence is
	characterised by the same algebraic power-law exponent, indicating a strong phase locking between collective excitations in all modes.
	Interestingly, such a phase-locking takes place also  {in} the vicinity of the critical point, 
	where strong phase fluctuations are expected to modify the mean-field picture \cite{dagvadorj2015nonequilibrium}.
	In contrast, we discover that, unlike the collective excitations (the sound modes), the topological defects (the vortices) in different components are not strongly correlated close to the phase transition, where the density of the fluid is small, nor further from the transition for the component with the weakest density i.e. the photonic idler mode.  
	Remarkably, the photonic idler component is found to possess an algebraic order, characteristic of two-dimensional superfluids, in the presence of {multi-vortex configurations}. 
	This is possible  only due to the coupling of the low-density idler mode to higher density components. 
	To our knowledge, such a peculiar state has never been seen before.
	This raises questions about the importance of the interplay between sound modes and vortices in superfluid phase transitions.		

	\paragraph*{ {System and theoretical modelling. ---}}

\begin{figure}[t]
		\begin{center}
			\includegraphics[width=\columnwidth]{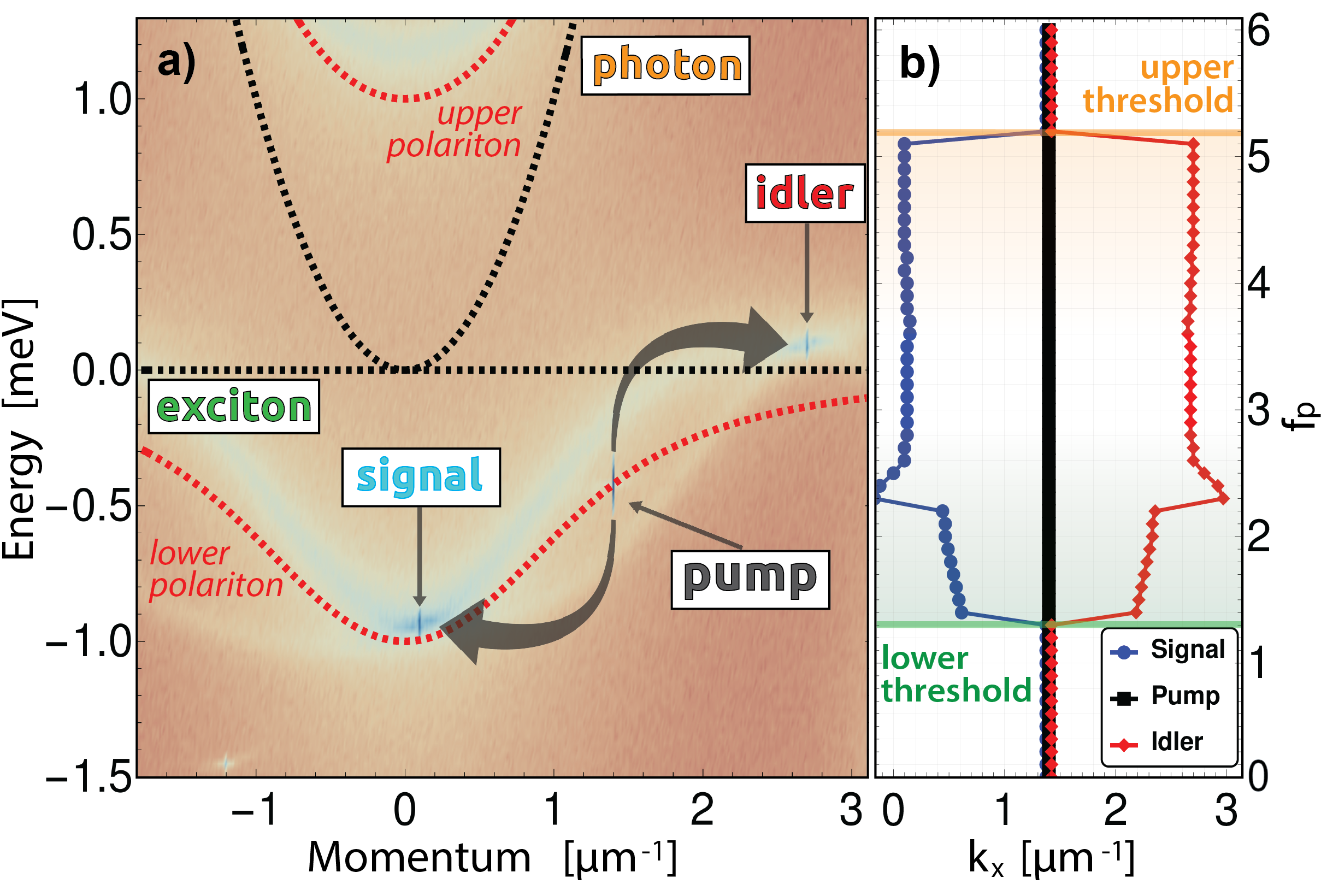}
		\end{center}
		\caption{\textbf{Polariton OPO regime.}
			\textbf{(a)} Typical spectrum of the system within the OPO window showing signal, pump and idler modes on the lower polariton branch. 			%
			\textbf{(b)} Signal, pump and idler momentum $k_x$ as a function of the pump strength $f_p$, exhibiting a double-threshold. 
			The ordered (disordered) phase lies in the pump range within (outside) a lower (solid green) and an upper (solid orange line) critical point. 
		}
		\label{Fig1}
	\end{figure}

We use stochastic simulations based on the truncated Wigner approximation~\cite{carusotto2005spontaneous}. Our method considers the full two dimensional multimode polariton field,  which includes fluctuations in both density and phase, represented by  complex number  {fields}  $\psi_\chi (\vect{r},t)$, where $\chi=\mathrm{X,C}$ denotes the single excitonic and photonic component, respectively. The stochastic differential equation for trajectories of  {these fields} takes the following form \cite{carusotto2013quantum,carusotto2005spontaneous,dagvadorj2015nonequilibrium} ($\hbar=1$)
	\begin{equation}
		i \text{d} \begin{pmatrix} \psi_{\text{X}} \\
			\psi_{\text{C}}\end{pmatrix} = \left[\hat{H}_0 \begin{pmatrix} \psi_{\text{X}} \\
			\psi_{\text{C}}\end{pmatrix}  + \begin{pmatrix} 0\\ F \end{pmatrix}\right]
		\text{d}t\\
		+ i \begin{pmatrix} \sqrt{\kappa_{\text{X}}} \text{d}W_{\text{X}}^{} \\
			\sqrt{\kappa_{\text{C}}} \text{d}W_{\text{C}}^{}\end{pmatrix} \; .
		\label{eq:sgpeq}
	\end{equation}
	In this notation, the Hamiltonian operator
	\begin{equation}
		\hat{H}_0 = \begin{pmatrix}  - \frac{\nabla^2}{2 m_{\text{X}}} - g_{\text{X}}|\psi_{\text{X}}|_-^2 
			-i \kappa_{\text{X}} & \Omega_{\text{R}}/2 \\ 
			\Omega_{\text{R}}/2 & - \frac{\nabla^2}{2 m_{\text{C}}}
			-i \kappa_{\text{C}} \end{pmatrix} \; ,
		\label{eq:sgpeq2}
	\end{equation}
	is written in terms of 
	the Rabi splitting $\Omega_{\text{R}}$, the exciton-exciton interaction strength $g_{\text{X}}$,
	the mass  $m_{\chi}$ and  the damping rates $\kappa_{\chi}$, where $\chi=$ X,C indicates excitons and photons, respectively.
	Here, we consider a homogeneous monochromatic continuous-wave pump $F(\vect{r},t) = f_p \exp{i (\vect{k}_p \cdot \vect{r} - \omega_p t)}$ with momentum $\vect{k}_p$, strength $f_p$ and frequency
	$\omega_p$, resonant with the bare  {lower}-polariton dispersion, so that, as shown in Fig.~\ref{Fig1}(a), polaritons undergo parametric scattering into the signal and idler states \cite{ciutiOPOtheory} for  {a} certain range of pump strengths $f_p$.
	$\text{d}W_{\chi}^{}$
	are independent white complex Gaussian noise terms, with zero mean and
	local correlations in time and space: 
	$\langle \text{d}W_{\chi}^{} (\vect{r} , t) \text{d}W_{\chi'}^{*} (\vect{r} , t) \rangle = {2} \delta_{\chi,\chi'} {\delta_{\vect{r},\vect{r}'}} \text{d}t /{\text{d}A}$,  
	and $\langle \text{d}W_{\chi}^{} (\vect{r} , t) \text{d}W_{\chi'}^{} (\vect{r} , t)\rangle = 0$. 
	The reduced Wigner density reads $|\psi_{\text{X}}|_-^2 = \left(|\psi_{\text{X}}|^2 - 1/dV\right)$.
	Physical observables  can be calculated by appropriate averages over stochastic realisations.

	For the numerics, we have considered specific system parameters, relevant for current
	experiments: $\Omega_{\text{R}} = 4.4$~meV, $m_{\text{C}} = 2.3\times 10^{-5} m_0$, where $m_0$ denotes the 
	electron mass,  $\kappa_{\chi} = 0.1$~meV, and
	$g_{\text{X}} = 2 \times 10^{-3}$~meV$\mu$m$^{2}$.
	We have ignored the exciton dispersion as $m_{\text{X}} \gg
	m_{\text{C}}$. 
	Eqs.~\eqref{eq:sgpeq} are numerically  implemented on a  $N^2$ grid with $N=256$ points and length $L = N \times a$,
	where $a = 0.87 \mathrm{\mu m}$ is the  {uniform grid} spacing. The pump is injected at finite momentum $\vect{k}_p = (k_p, 0)$ in the $x$-direction, with $k_p = 1.6$~$\mu$m$^{-1}$.	
		
	We evolve the stochastic equations~\eqref{eq:sgpeq} in a two-dimensional geometry with periodic boundary conditions until the steady-state.
	While mean-field wavefunctions are used as initial conditions, the Weiner noise terms are adiabatically switched  {on} along the dynamics.
	As explained in Ref.~\cite{SM}, once the steady-state is achieved, for each $\chi$-component a filtering process allows  {us} to extract the different fields $\psi_{\chi,n} (\vect{r},t)$, and the momenta $\textbf{k}_{\chi,n}$ at which they are peaked in the momentum distribution, where $n={s,p,i}$ labels the \textit{signal}, \textit{pump} and \textit{idler} mode, respectively. We ascertain that at a given pump value, the set of modes $n$ have the same momentum structure in both components $\chi$~\cite{SM}. 
	In Fig.~\ref{Fig1}(b) we plot the behaviour of the extracted momenta $k_x$ for the photonic component, where $\textbf{k}_{\text{C},n}  = (k_x,0)$, as a function of the pump strength, for each mode $n$.
	Note that the OPO quasi-ordered phase possesses  two different thresholds, easily distinguishable from Fig.~\ref{Fig1}; we extract a lower threshold (LT) at $f_p = 1.419$ and an upper threshold (UT) at $f_p=5.149$.
	Further details on the filtering process, the resulting steady-state diagram and identification of the critical points are reported in Ref.~\cite{SM}.
	
	As discussed earlier, at  {the} mean-field level the sum of the OPO signal and idler phases is locked to that of the external pump $F$. 
	In the next sections we show how fluctuations can alter such phase locking~\cite{dagvadorj2015nonequilibrium}, and its implications on the observation of unconventional order in a multicomponent phase transition.
		
	\paragraph*{ {Onset of multicomponent order. ---} }
	\begin{figure}[t]
		\begin{center}
			\includegraphics[width=\columnwidth]{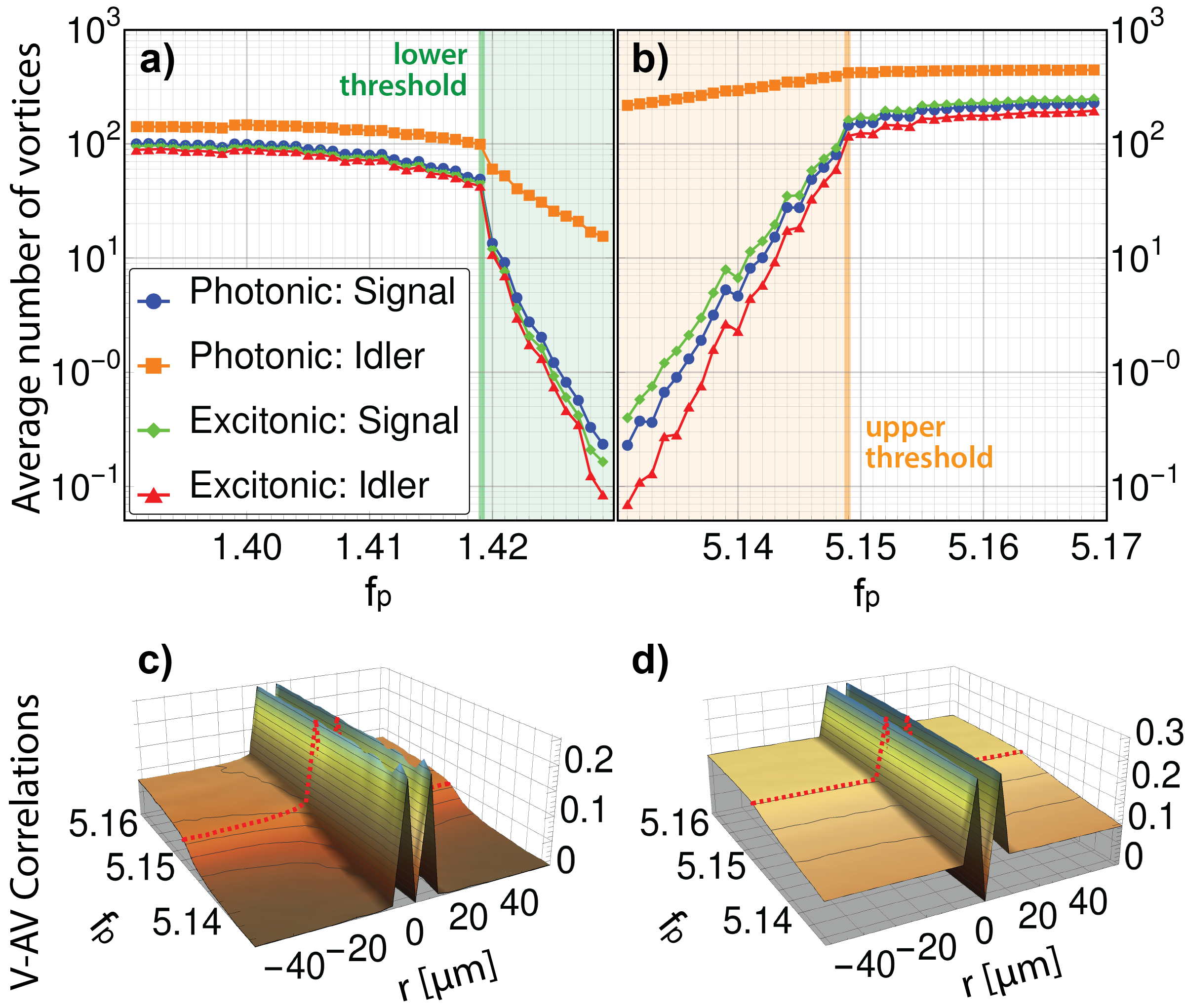}
		\end{center}	
		\caption{
			\textbf{Vortices near the OPO thresholds.}
			\textbf{Top:}
			averaged number of vortices/antivortices in all four components, around the (a) lower  and (b) upper thresholds. 
			The quasi-condensate phase is shown as a shaded coloured area.
			\textbf{Bottom:} 
			Vortex-antivortex (V-AV) correlations at the upper threshold for the photonic component in the signal (c) and the idler (d) modes. 
		}
		\label{Fig2}
	\end{figure}
	
\begin{figure}[t]
		\begin{center}
			\includegraphics[width=0.5\textwidth]{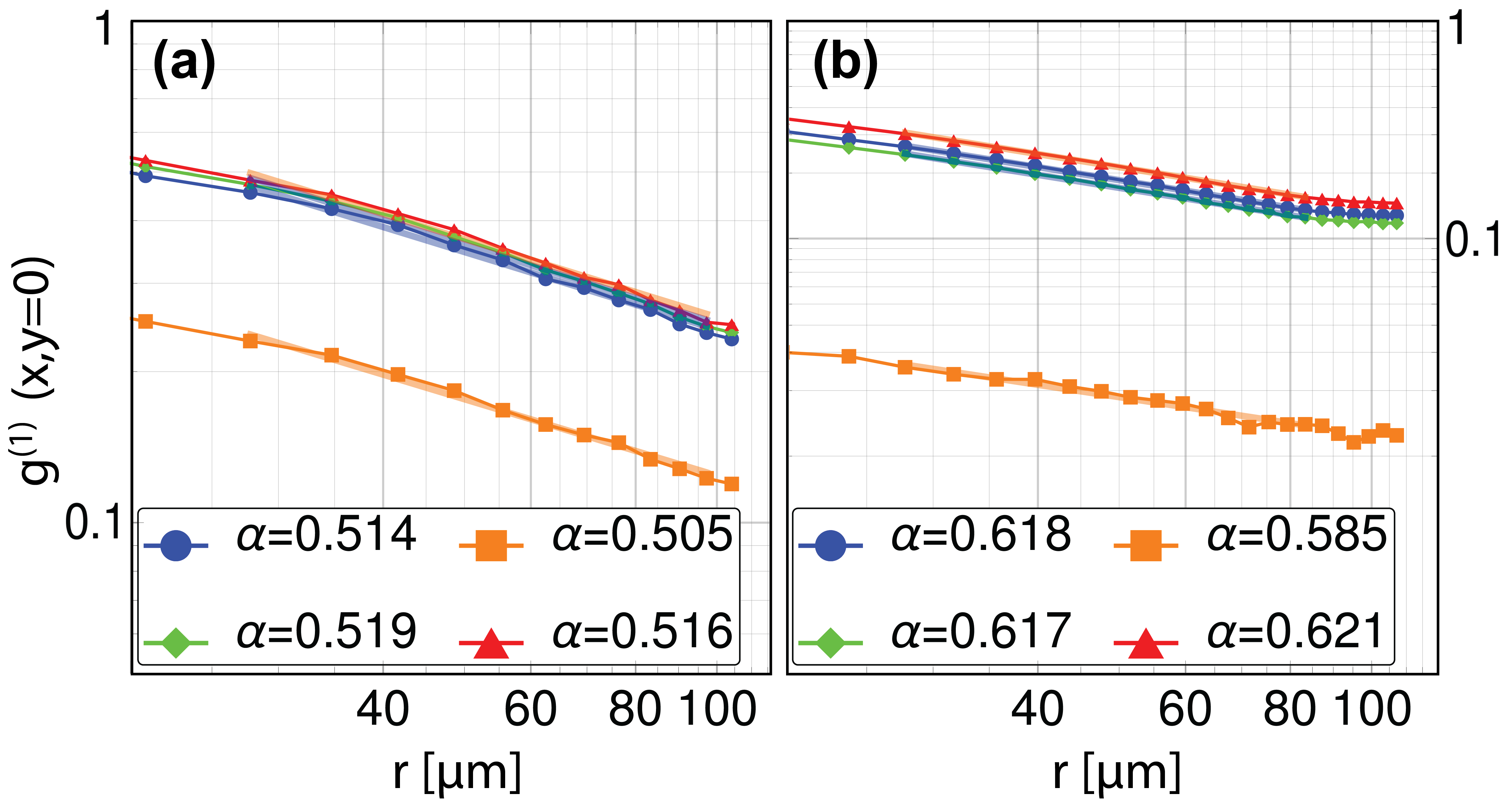}
			\includegraphics[width=0.5\textwidth]{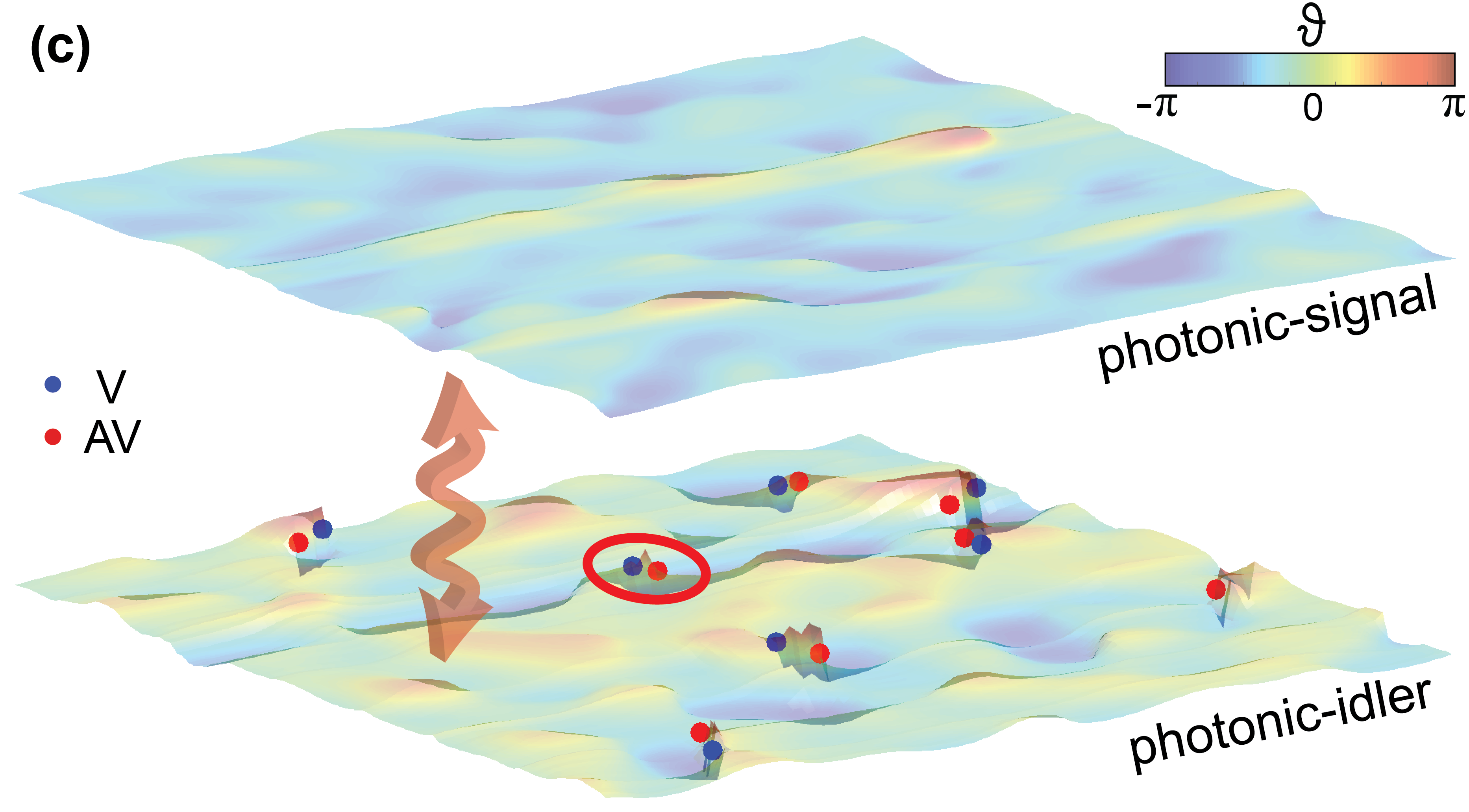}
		\end{center}
		\caption{\textbf{Inter-components phase-locking.}
			First order spatial correlation function along the $x-$direction at the lower $f_\mathrm{th}^\mathrm{low} = 1.419$ (a) and the upper $f_\mathrm{th}^\mathrm{up} = 5.149$ (b) threshold. The power-law exponents $\alpha$ (insets) show long-range phase-locking between collective modes in different components.
			(c)  {Zoomed-in view of} the real-space phase-profiles at the lower threshold. Note, the presence (absence) of vortex configurations in the photonic idler (signal) {, while away from the vortices the phase still exhibits correlation between the two components.} 
			}
		\label{Fig3}
	\end{figure}

	We proceed by investigating the nonequilibrium   phase diagram of our multicomponent system.
	As in previous works~\cite{dagvadorj2015nonequilibrium,comaron2018dynamical,Comaron2021}, we calculate the average number of vortices (Fig.~\ref{Fig2}) and spatial correlation functions (Fig.~\ref{Fig3}) in the steady-state close to the lower (LT) and upper (UT) phase transitions.
	We see that, for all components, while in the disordered phases, below (above) the LT (UT), the {total vortex/antivortex} number is large and almost constant. When the pump strength passes the critical point, the number of topological defects abruptly decrease. 
	This is in agreement with previous results\cite{comaron2018dynamical}: in the late dynamics, the system approaches a steady-state number of vortices, which becomes lower as one goes deeper into the ordered phase.

	While we observe an abrupt change in the steady-state number of vortices at the same pump power for all components, we note that they host a different number of vortices at a given pump power --- the inter-component phase locking is not perfect in the vortex channel beyond the mean-field level.
	In particular, the photonic idler mode  contains a number of topological defects which is, in the ordered phase, orders of magnitude larger than in the case of the other components.
	This  can be understood by noting the difference in the particle densities (reported in Fig.~1b of Ref.~\cite{SM}):
	at small densities, characteristic of  {the} photonic-idler, critical fluctuations are relatively stronger  and the creation of {random vortex-antivortex pairs is more} effective as compared to other higher density components.
	Such a behaviour is in line with the observations in Refs.~\cite{Comaron2021,Mei2021}.
	%
	
	The first order spatial correlation function in an isotropic system at a given time $t$, reads:
	\begin{equation}
	g^{(1)} (\vect{r}) = 
	\Frac{\langle \psi_{s}^* (\vect{r} + \vect{R})
		\psi_{s}^{} (\vect{R}) \rangle_{\vect{R},\mathcal{N}}}
	{\sqrt{ \langle \psi_{s}^*
			(\vect{R}) \psi_{s}^{} (\vect{R}) \rangle_{\vect{R},\mathcal{N}} \langle 
			\psi_{s}^*(\vect{r} + \vect{R}) \psi_{s}^{} (\vect{r} + \vect{R})
			\rangle_{\vect{R},\mathcal{N}}}}\; ,
	\label{eq:corre}
	\end{equation}
	where the average $\langle \cdots \rangle$ is performed over the spatial position $\textbf{R}$ and a large number $\mathcal{N}$ of stochastic realisations.
	Our numerics show that all components exhibit a phase transition  {from} exponential to power-law decay of correlations at the same pump power, similarly to the abrupt decrease of the number of topological defects analysed in Fig.~\ref{Fig2}. 
	Such a transition point has been quantitatively estimated by comparing  different fitting functions and choosing the best fit, as in the methods reported in Refs.~\cite{dagvadorj2015nonequilibrium,Comaron2021}.
	In Fig.~\ref{Fig3} we plot $g^{(1)} (\vect{r})$ \eqref{eq:corre} at the critical pump powers for the LT (a) and  UT (b). 
	Although the four modes contain very different  {numbers} of vortices, we extract ({within error bars}) the same algebraic exponents, characterising the long-range decay of $g^{(1)} (\vect{r})$  \eqref{eq:corre}.
	A detailed discussion is reported in Ref.~\cite{SM}.
	
	Motivated by the unusual observation of  {a} high density of defects in the algebraic phase of the photonic idler, even far from the critical point, we investigate further their spatial distributions. 
	{An example of a single-realisation of the photonic signal and photonic idler phases is shown in Fig.~\ref{Fig3}(c).}
	In the conventional BKT transition, away from the critical point on the quasi-ordered side, vortices are expected to be either fully paired or absent.
	On the contrary, the real space vortex configurations in the photonic idler, especially close to the upper threshold, show presence of 
	{
	 multi-vortex configurations even at pump powers deep in the ordered phase (see Figs.~{S9-S10} and discussion in Ref.~\cite{SM}). 
	}
	This visual observation is corroborated by the vortex-antivortex space correlations, reported in Fig.~\ref{Fig2}~(c,d) and discussed in more detail in the next section.

	\paragraph*{ {Vortex correlations and pairing in single component. ---}} 
	
	In order to gain a better understanding	of correlations  between vortices in different components, as well as the pairing of vortices and antivortices known to underly the BKT phase transition, we proceed by computing various vortex correlation functions.
	We define the spatial number correlators  between vortex-vortex/antivortex-antivortex (hereafter denotes as V-V), and vortex-antivortex/antivortex-vortex (V-AV) as:
	\begin{equation}
	\eta^{(1)}_{\{ \text{V:}\alpha\beta\gamma, \alpha'\beta'\gamma'\}} (\vect{r}) = \Frac{\langle n_\text{V}^{\alpha \beta \gamma}(\vect{r},t)n_\text{V}^{\alpha' \beta' \gamma'}(\vect{r}+\vect{R},t)\rangle_{\vect{R},{\mathcal{N}}}}{\sqrt{\langle N_\text{V}^{\alpha\beta\gamma} \rangle_{\mathcal{N}}\langle N_\text{V}^{\alpha'\beta'\gamma'} \rangle_{\mathcal{N}} }},
	\label{eq:vortex_corr}
	\end{equation}
	where $n_\text{V}(\vect{r})$ and $\langle N_\text{V} \rangle$ are the spatial distribution and average number of vortices with components $\alpha\in\{\text{Vrtx},\text{AntV}\}, \quad \beta\in\{\text{Sig},\text{Idl}\} \quad \text{and} \quad \gamma\in\{\text{Ph},\text{Ex}\}.$
	
 	Let us first focus on V-AV correlations.
 	An example is shown in panel (c) and (d) of Fig.~\ref{Fig2}, where the 3D surfaces show the correlator  \eqref{eq:vortex_corr} as a function of the relative distance $r$ as the pump-power is varied across the phase transition.
 	The threshold point is  {indicated} as a dashed red curve on top of the 3D surface.
	In all cases, the correlator assumes {a} double peak structure with a minimum at the origin indicating V-AV attraction leading to annihilation events. 
	{ 	In the disordered phase, characterised by  {a} high number of free vortices,
 	we find that  V-AV correlations  {asymptotically approach} a  finite (non-zero) value as $r \to \infty$, indicating presence of vortices, whose binding nature is undetermined.		
	In the ordered phase of the photonic signal (Fig.~\ref{Fig2} c), V-AV correlations plateau to a null value, indicating that on average only one tightly-bound V-AV pair is left in the system.
 	On the other hand, the ordered phase of the photonic idler (Fig.~\ref{Fig2}(d)), in agreement with Fig.~\ref{Fig2}(a), still contains a large number of vortices as the correlator plateaus to a smaller but finite value. 
 	By visually inspecting  {the} spatial distribution of vortices in Figs.~{S9-S10} of Ref.~\cite{SM}, we find multi-vortex configurations. 
 	Note, that this ensemble of vortices is mainly composed of vortex pairs; interestingly however, we also find indication of the presence of rare free vortices.}
 	Furthermore, in Figs.~{S5-S6} of Ref.~{\cite{SM}} we investigate and compare V-AV correlations in different components.	We find that all components exhibit the same pairing crossover from a ``randomly distributed" disordered phase to a ``{single-pair}" ordered phase, except for the photonic-idler.
	For pump strengths both above LT and  below UT,  we find a non-zero plateau indicating presence of {a high number of vortices}.
	The UT case is depicted in Fig.~\ref{Fig2}(d).

	We also compute V-V correlations, which are shown and discussed in Figs.~{S2}- {S4} of Ref.~{\cite{SM}}. Similarly to  {the} V-AV case, for all components other than the idler, we find a finite-valued plateau at large distances ({many} vortices), which crossovers to zero ({single-pair}) as one approaches the ordered phase. 
 	Close to the origin the autocorrelations drop to zero, indicating that vortices repel each other.
 	Again, similarly to the V-AV, the V-V correlations  {also} show that the photonic-idler component, even deep in the ordered phase, still shows a (lower than the one in the disordered phase, but still noticeable) {finite-valued plateau}.

	\begin{figure}[]
		\begin{center}
			\includegraphics[width=\columnwidth]{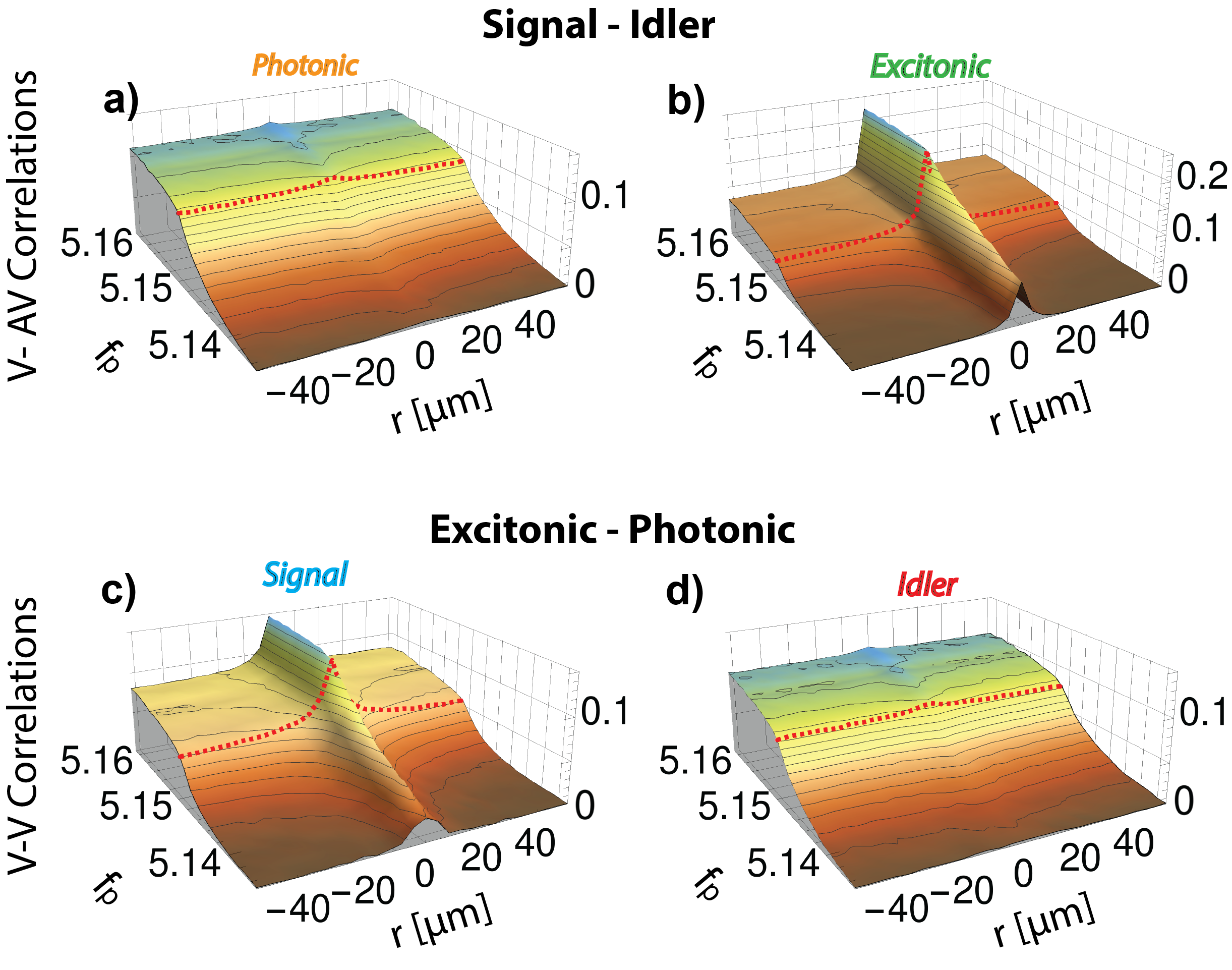}
		\end{center}
		\caption{
			\textbf{Correlations between vortices in different components at the upper threshold.}
			\textbf{Top:} Correlation between vortices in signal and anti-vortices in idler for  the photonic (a) and the excitonic (b) component.
			\textbf{Bottom:} Correlation between vortices in photonic and excitonic components for signal (c) and idler (d) modes.}
		\label{Fig4}
	\end{figure}
	
	\paragraph*{ {Correlations between vortices in different components. ---}}
		Considering the processes underlying the symmetry-breaking mechanism of parametric regimes
	---~corresponding to an opposite rotation of the phases of  {the} signal and idler components~---
	it is expected that vortices in different modes would be ``position-locked" while possessing opposite circulation~{\cite{carusotto2005spontaneous}}. 
	Vice-versa, when considering vortices in the same mode (signal or idler) but different component (excitonic and photonic), 
	the locking would take place between two vortices with same sign-circulation.
	Following this reasoning, it is straightforward  to chose  {the} V-AV (V-V) correlations to quantify same-component (same-mode) vortex-locking.

	First, let us discuss the former case: signal-idler V-AV correlations are shown in Fig.~\ref{Fig4}(a),(b) for  {the} photonic and excitonic components, respectively.
	Here, vortex-locking is easily quantified by how smooth (or peaked) the V-AV correlation profiles are across the transition: 
	while a homogeneous profile along the $r$-axis would indicate a similar probability of finding the V or AV partner everywhere,
	a peaked distribution would rather show that it is more probable to find such objects in the same position.
	Results shown in Fig.~\ref{Fig4}(a),(b) reveal that  {this inter}-mode spatial pairing mechanism takes place only for the excitonic fraction, but not for the photonic one.

	A similar analysis is carried out for investigating the single-mode vortex-coupling between  {the} excitonic and photonic components.
	The V-V correlator is shown in Fig.~\ref{Fig4}(c) for the signal and in Fig.~\ref{Fig4}(d) for the idler modes.
	The results reveal a strongly-peaked distribution for the signal case, while the idler mode shows a smooth profile.
	{
	We conclude that the spatial vortex-locking between components and/or modes is always strong apart from the cases with photonic-idler mode;
	this is a further confirmation of a weak vortex-locking mechanism observed for the photonic-idler component.
	}

	\begin{figure}
		\begin{center}
			\includegraphics[width=0.5\textwidth]{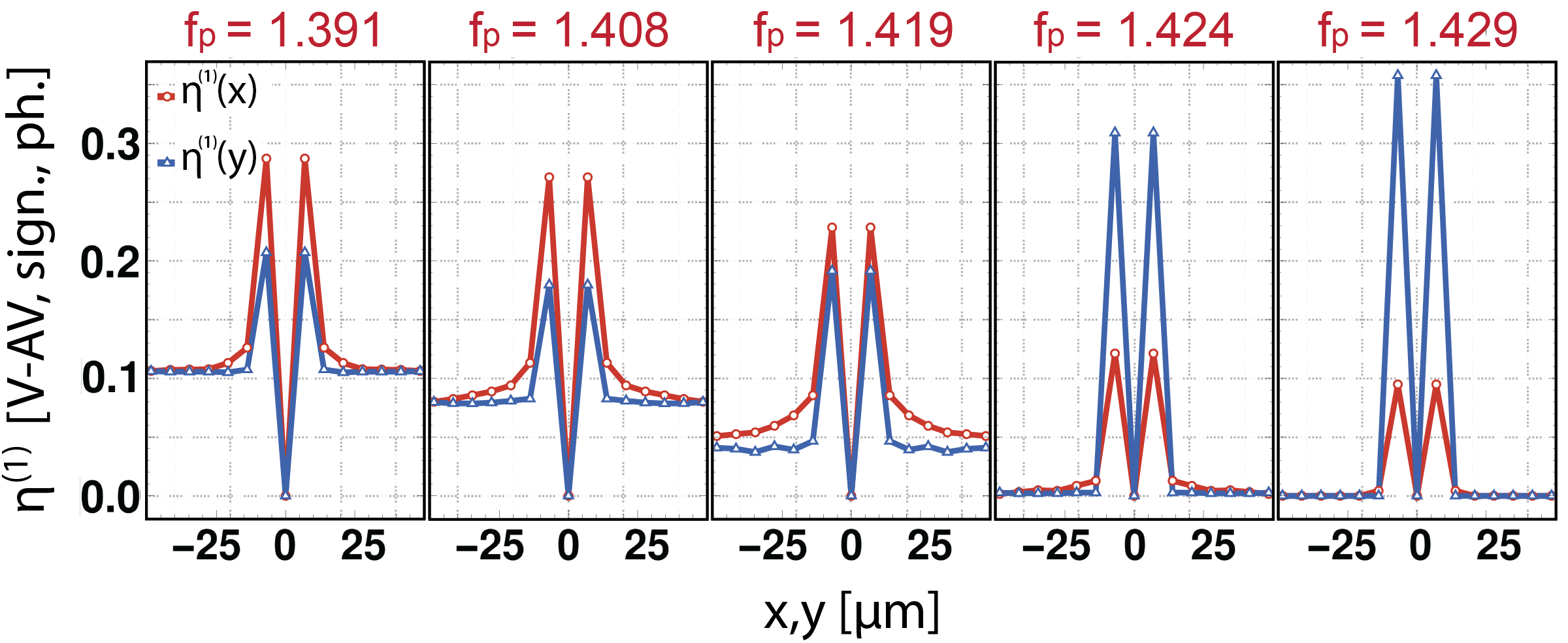}
		\end{center}
		\caption{\textbf{Correlations between vortices and antivortices in the signal photonic component} along $x$ (red curves) and $y$ (blue curves) directions for different pump powers across the lower threshold.
		}
		\label{Fig5}
	\end{figure}

	\paragraph*{ {Vortex-antivortex correlation anisotropy. ---}}
	Finally, we discuss the presence of anisotropic features in the pairing mechanism, which we observe in the vortex correlation analysis discussed above.
	Let us focus on the signal mode in the photonic component near the two thresholds. 
	Fig.~\ref{Fig5} shows the V-AV correlation function in the photonic signal across the LT, along two orthogonal directions $x,y$.
	Here we see that the magnitude of pairing is different along the x and y directions. Moreover, the dominant pairing direction  switches across the BKT phase transition. Near the UT, vortex correlations are instead isotropic (see Fig.~{S6} in Ref.~\cite{SM}).
	To explain this behaviour, we consider the possibility for the V-AV pairs to feel anisotropic forces, possibly inherited from the anisotropic pump, which is injected in the $x-$direction.
	We proceed by checking the diffusion coefficients of the quantum fluid, which can be extracted by mapping the exciton-polariton model to the Kardar–Parisi–Zhang (KPZ) equation~\cite{KPZ1986} for the phase. Note, that analytical estimates for the vortex-vortex interactions in this problem exist only for the isotropic case \cite{wachtel2016electrodynamic} allowing only approximate analysis.
	The results obtained are not conclusive (see a detailed discussion in Ref.~\cite{SM}), leaving this as an exciting  open question to be addressed in future works. 

	\paragraph*{ {Conclusions. ---}}
	
	{In this work, we have  investigated theoretically the multicomponent BKT phase transition of a two-dimensional driven-dissipative  {bosonic} system in the OPO regime.
	Clear signatures of the transition from a disordered to an ordered state are inferred by computing the first order coherence, the number of vortices and vortex correlations.
	We find, that the critical point occurs at the same pump power with the same exponent of the algebraic decay of the first order spatial coherence function for all components.
	However, collective phase fluctuations (the sound modes) and the vortices are found to behave quite differently: while collective excitations are locked across all components, vortices are not perfectly spatially correlated, especially those involving the photonic idler {at the upper threshold}. 
	This results {in a new phase, where an algebraically ordered state exists in  {the} presence of multi-vortex configurations, 
	 composed of V-AV pairs and occasionally free vortices}.
	Such a peculiar phase is, to our knowledge, unique in the realm of condensed matter systems, and sets the basis for a novel form of superfluidity, where a low density quantum fluid is populated by
	{topological defects}, but remains in the algebraically ordered phase due to locking  {of} the collective excitation channel to other higher density components.}

	\paragraph*{ {Acknowledgements. ---}}
	We thank I.~Carusotto, F. M. Marchetti and A. Zamora for stimulating discussions. We thank A. Ferrier for proofreading the manuscript.
	MHS, GD and PC gratefully acknowledge financial support from QuantERA InterPol and EPSRC (Grant No. EP/S019669/1, EP/R04399X/1 and No. EP/K003623/2).
	%
	
%

\end{document}


	
	\title{Supplementary material for: Unconventional Berezinskii-Kosterlitz-Thouless Transition in the Multicomponent Polariton System}

	\author{G. Dagvadorj} 
	\affiliation{Department of Physics, University of Warwick, Coventry,
		CV4 7AL, UK}
	
	\author{P. Comaron}
	\email[Corresponding author: ]{p.comaron@ucl.ac.uk} 
	\affiliation{Institute of Physics, Polish Academy of Sciences, Al. Lotnik\'ow 32/46, 02-668 Warsaw, Poland}
	\affiliation{Department of Physics and Astronomy, University College
		London, Gower Street, London, WC1E 6BT, UK}
	
	\author{M.~H.~Szyma\'nska} 
	\affiliation{Department of Physics and Astronomy, University College
		London, Gower Street, London, WC1E 6BT, UK}

	
	\maketitle
	
	\section{Numerical simulations and methods}
	\begin{figure*}
		\centering
		\includegraphics[width=0.4\textwidth]{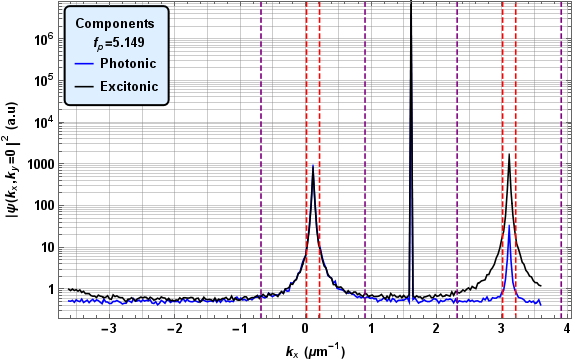}
		\includegraphics[width=0.4\textwidth]{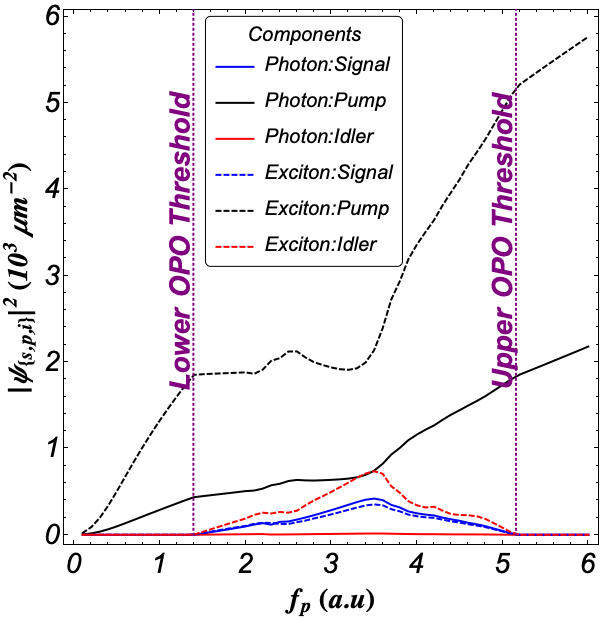}
		\includegraphics[width=0.4\textwidth]{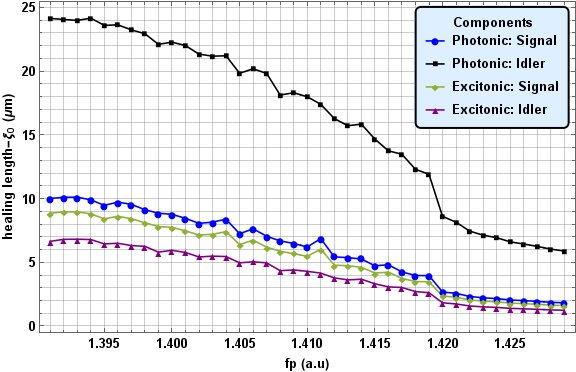}
		\includegraphics[width=0.4\textwidth]{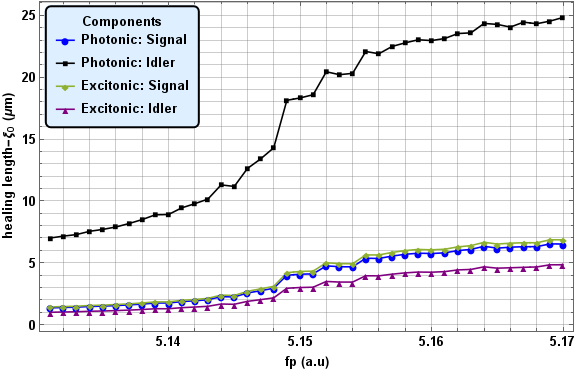}		
		\caption{
			\textbf{Top left:} Density momentum distribution along the $k_x$ axis ($k_y=0$) of the stead-state density field at the pump value $f_p = 5.149$ corresponding to the upper OPO threshold.
			A filtering procedure is applied to extract the single-mode wavefunction and the healing length; boundaries of the selected filtering windows are shown as purple and red vertical dashed lines, respectively.
			\textbf{Top right:} The average density phase diagram  as a function of the pump strength for each OPO mode and exciton/photon component. 			\textbf{Bottom:} The healing length from Eq.~\eqref{eq:healinglength} as a function of the pump strength for the different components in intervals around the lower (left panel) and upper (right panel) OPO thresholds.
		}
		\label{Fig1_SM}
	\end{figure*}

	In this work, we numerically study the phase transition in a two-dimensional microcavity under the optical parametric oscillator (OPO) regime~\cite{carusotto2013quantum}.
	The approaches we use to obtain the polariton phase diagram are inspired by the methods presented in our recent works~\cite{dagvadorj2015nonequilibrium,Comaron2021}, and explained in the main text.
	In this section, we give a detailed explanation of the procedure we adopt to extract the steady-state of the system.
	As explained in detail in the literature, investigation of the phase transition can be undertaken at a mean-field (MF) or above mean-field level, for instance, adopting Truncated-Wigner (TW) methods~\cite{carusotto2013quantum}.
	Our analysis is comprehensive of both pictures. 
	
	For the latter case, the numerical steady-state configurations of the system at each pump value discussed in the main text are obtained by simulating the equations of motion of the polariton system (i.e. Eq.~\red{(1)} in the main text) 
	using the numerical MF steady-states as initial conditions for the corresponding pump powers. 
	The nonequilibrium steady state is reached after each realisation is converged; then, stochastic averages are applied to obtain the physical observables investigated (such as the density field, coherence and number of topological defects).
	Each steady state is the result of averaging over $\mathcal{N}$ different stochastic realizations.
	We note that the final configuration is independent of the chosen initial conditions: the final steady states can be also reached starting from an initial noise configuration and then applying a sudden quench for the given driving~\cite{dagvadorj2015nonequilibrium}.

	\subsection{Identification of the OPO components and filtering procedure for the single-component wavefunction}
	\label{sec:filtering}
	Once the steady state for a single pump power is computed, we proceed by identifying the primary three modes of the system, i.e. pump, signal and idler. 
	The position of each component is extracted from the whole field through a momentum-selection procedure.
	Such a procedure is important in order to filter the fundamental components out of the whole field, which can also exhibit the presence of satellite modes at different momenta. 
	
	In order to locate the signal and idler momenta we proceed as follows. 
	First, we numerically calculate, for each component, the density distribution in momentum space from the steady-state real-space wavefunction.
	Then, we select the momentum density distribution for $k_y=0$. 
	Finally, the signal and idler modes are identified by the conditions
	\begin{eqnarray}
	k_s&=&\max[|\psi(k_x,k_y=0)|^2]_{k_x<k_p} \\
	k_i&=&\max[|\psi(k_x,k_y=0)|^2]_{k_x>k_p},
	\label{eq:filtering}
	\end{eqnarray}  
	where $k_p$ is the momentum associated with the pump mode, as introduced in Eq.~\red{(1)} of the main text.
	We find that this condition is sufficient to locate the modes, even for those cases where the momentum distributions are found to be very noisy. 
	It is also worth noting that satellite states in our results tend to populate the field at pump values which are away from the critical points;
	this is due to the enhanced stability of the primary modes around the thresholds, which suppresses the scattering of particles into satellite modes, in agreement with the results discussed in Ref.~\cite{dunnett2018properties}.

	In the upper left panel of Fig.~\ref{Fig1_SM}, we show an example of the steady-state density distribution in $(k_x,k_y=0)$-space at a given pump power; here, the whole photonic (excitonic) field at (what we will later find to be) the upper threshold, is shown as solid blue (black) line.
	Once the momenta of these primary modes have been located, in order to obtain the corresponding single-mode wavefunctions,
	we proceed by applying a \textit{filtering process} that allows us to not only filter out spurious satellite states, but also control the intake of noisy fluctuations in the field. 
	The filtering consists of selecting only some specific modes that lie within an interval centred around the momenta $k_s$, $k_p$ and $k_i$.	
	For the upper-threshold case shown in the upper left panel of Fig.~\ref{Fig1_SM}, the different filtering regions around the specific momenta are identified as vertical dashed purple lines.
	Such regions are symmetric with respect to the primary mode momenta and have been chosen so that one of the boundaries lies exactly in the middle of two of the located momenta.
	Finally, by transforming back from momentum to real space, we are then able to obtain the filtered wavefunction for each mode of the specific component, wavefunctions that we will later use to calculate observables such as first-order correlations for the vortices and phase, and for identifying topological defects, important for the analysis discussed in the main text.
	Notably, this procedure can be implemented at both a MF and TW level.
	It is also important to note that the filtering procedure results in a modification of the lattice discretisation when transforming the field back from momentum-space to real space; this is an immediate consequence of the narrowing of the momentum window following the filtering, $k_\mathrm{max}^\prime = \pi/a^\prime < k_\mathrm{max}$, with $k_\mathrm{max}^\prime$ the maximum momentum and $a^\prime$ the grid discretization after the filtering process.
	Specifically, for our simulations we find the new lattice spacing to be $a^\prime \approx 6.944 \mu m$ at the lower and $a^\prime \approx 3.968 \mu m$ at the upper OPO thresholds.    

	\subsection{The OPO density phase diagram and location of quasi-condensation thresholds}
	After the positions of the signal, idler and pump momenta across the OPO region have been extracted, together with their filtered wavefunctions
	we proceed in discussing the methods used to compose the OPO density phase diagram and numerically locate lower and upper MF and TW thresholds.
	Such a procedure has been demonstrated to be nontrivial at the TW level \red{\cite{Comaron2021,dagvadorj2015nonequilibrium}}.
	We tackle this problem by extracting the densities of each component and mode, for different pump strengths.
	In order to obtain the densities of the signal and idler modes within both the MF and TW approaches, 
	we first calculate the photonic and excitonic component at different pump powers.
	Then, for each pump strength we apply the filtering process discussed in Sec.~\ref{sec:filtering}, allowing us to precisely extract the momenta of the primary modes; from the wavefunctions extracted, we can finally build up a \textit{density phase diagram} as a function of the driving intensity.

	Let us first focus on the TW case.
	From the results of the numerical integration of the full model \red{(1)} of the main text,
	by applying the filtering selection (upper left panel of Fig.~\ref{Fig1_SM}), 
	we are able to extract the average density of each primary mode of the excitonic and photonic fields, as a function of the driving strength (see Fig.~\ref{Fig1_SM}, upper right panel). 
	It is interesting to note that while the intensity of the pump mode in both components increases as the driving parameter is increased, the signal and idler components exhibit a double phase transition, where the symmetry-broken phase, where the condensate density is non-zero, sits between two distinct critical points. 
	These two thresholds are located at a low and a high pump regime, respectively, and reported in the upper right panel of Fig.~\ref{Fig1_SM} as dotted purple vertical lines.	
	
	In order to locate the quasi-order thresholds, it is instructive to investigate the behaviour of topological defects and the system's healing length in the non-equilibrium steady state across the phase transition.
	Thus, we proceed in detecting and counting the average number of vortices (for numerical methods, see discussions reported detail in our previous works~\cite{dagvadorj2015nonequilibrium,comaron2018dynamical}), and calculating the healing length $\xi_0$ for each pump power, which reads
	\begin{equation}
	\xi_0=\hbar/\sqrt{2mg_X|\psi_{s,i}^{X,C}|^2_{\Delta k < 0.1 \mu m^{-1}}}.
	\label{eq:healinglength}
	\end{equation}
	The healing length as a function of the driving strength is plotted in the lower panels of fig.~\ref{Fig1_SM}, for the lower (left panel) and upper (right panel) thresholds.
	To extract this length, we decide to adopt a more strict filtering condition around the signal and idler modes, i.e. $\Delta k < 0.1 \mu m^{-1}$. 
	This increases the precision of our results. In the upper left panel Fig.~\ref{Fig1_SM}, these integration regions are delimited by the red dashed lines.
	In agreement with previous numerical works~\cite{dagvadorj2015nonequilibrium,comaron2018dynamical,Comaron2021}, from the decay of the healing length and the number of vortices (the latter being discussed and shown in Fig.~\red{1} of the main text), 
	one can obtain a very good estimate for the upper and lower BKT thresholds within the TW approximation, which read:
	\begin{equation}
	f_p^{low-BKT}=1.419 \quad \text{and} \quad f_p^{upp-BKT}=5.149.
	\label{eq:}
	\end{equation}

	It is interesting to repeat the first part of this analysis (i.e. filtering process and density integration) within the MF-approximation, where we identify the critical points at $f_p = 1.391$ and $5.156$ for the lower and upper thresholds, respectively.
	
	\

	\begin{figure*}
		\begin{center}
			\includegraphics[width=0.4\textwidth]{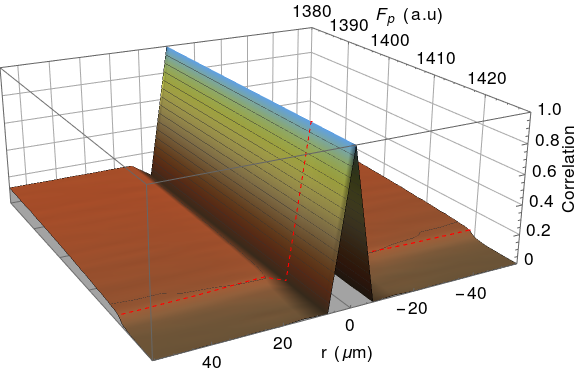}
			\includegraphics[width=0.4\textwidth]{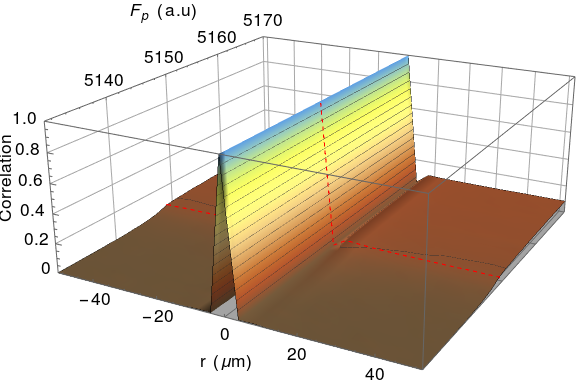}
			\includegraphics[width=0.4\textwidth]{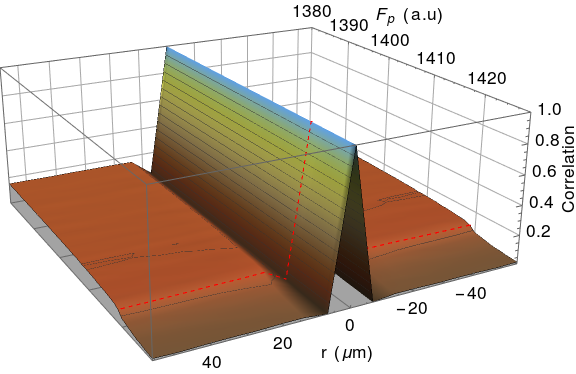}
			\includegraphics[width=0.4\textwidth]{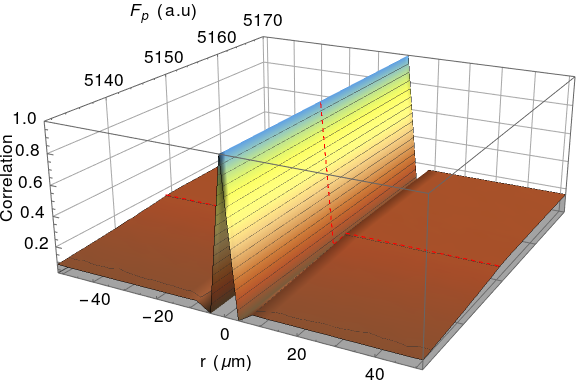}
		\end{center}
		\caption{
			Correlations between vortex-vortex/antivortex-antivortex in the same mode, for the photonic component. 
			Correlations in the signal (idler) mode are depicted in the top (bottom) row, respectively, 
			in the vicinity of the lower (left column) and upper (right column) OPO thresholds.
			The exact threshold point is highlighted as a dashed red line.
			}
		\label{figSM:2}
	\end{figure*}
		\begin{figure*}
			\begin{center}
				\includegraphics[width=0.4\textwidth]{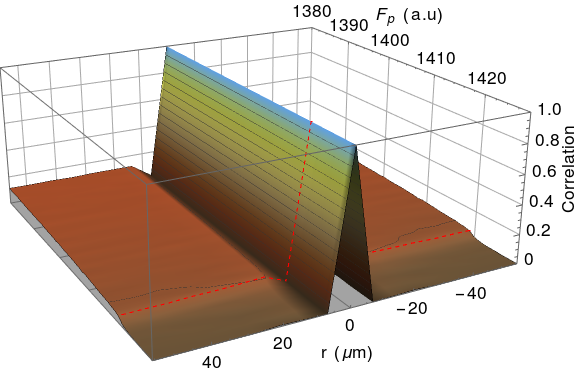}
				\includegraphics[width=0.4\textwidth]{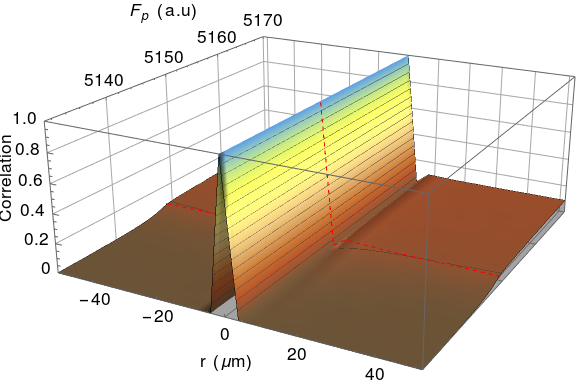}
				\includegraphics[width=0.4\textwidth]{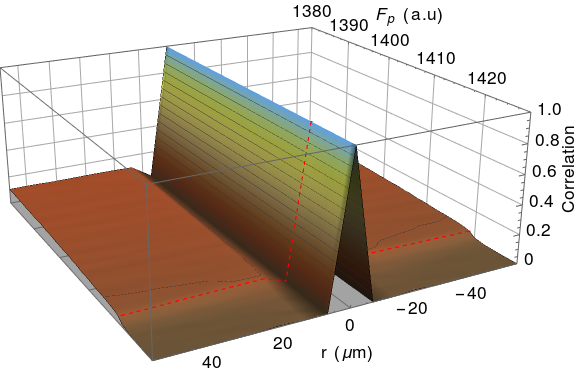}
				\includegraphics[width=0.4\textwidth]{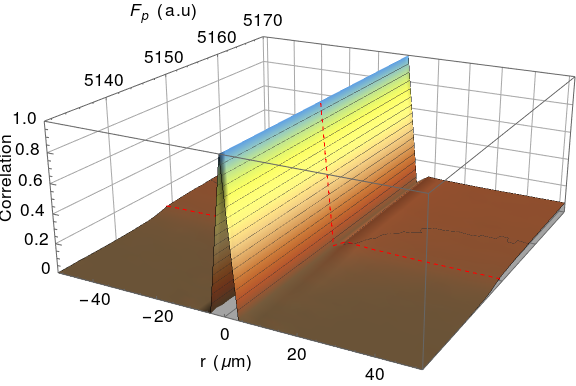}
			\end{center}
			\caption{As in Fig.~\ref{figSM:2}, but for the excitonic component.}
			\label{figSM:3}
		\end{figure*}	
	\begin{figure*} 
		\begin{center}
			\includegraphics[width=0.35\textwidth]{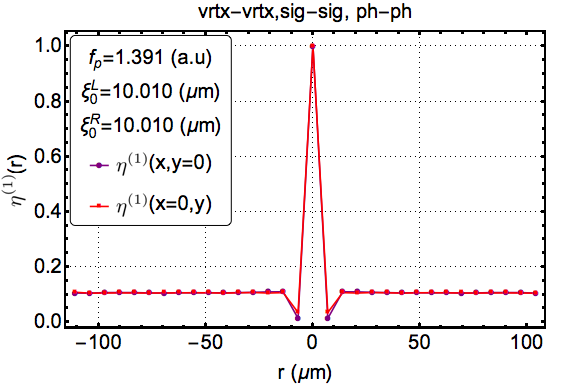}
			\includegraphics[width=0.35\textwidth]{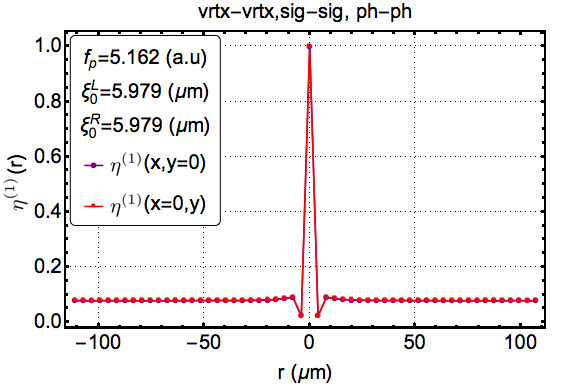}
			\includegraphics[width=0.35\textwidth]{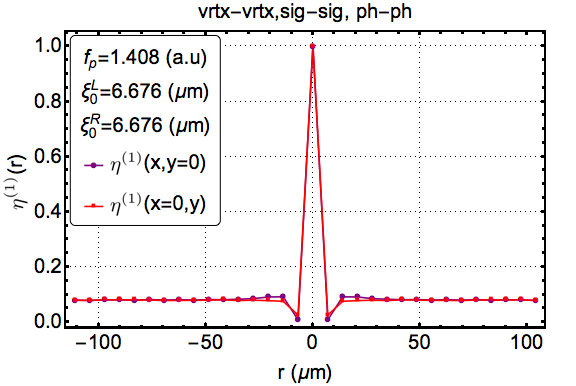}
			\includegraphics[width=0.35\textwidth]{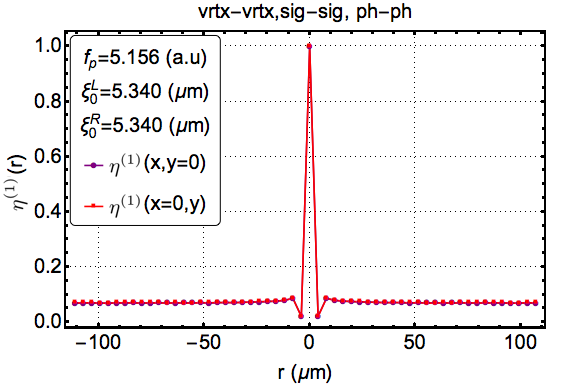}
			\includegraphics[width=0.35\textwidth]{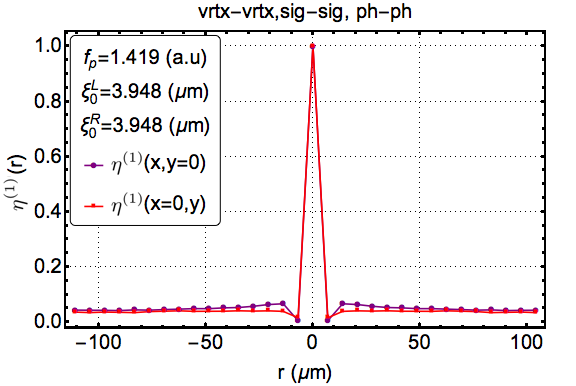}
			\includegraphics[width=0.35\textwidth]{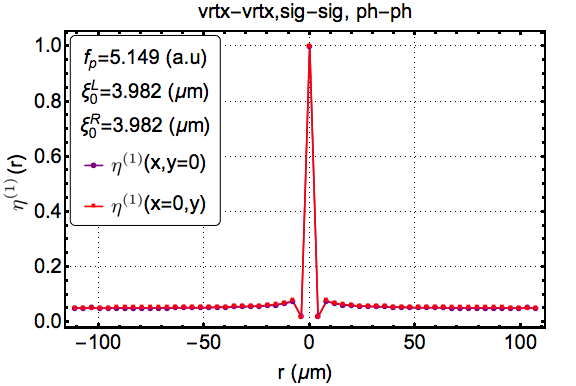}
			\includegraphics[width=0.35\textwidth]{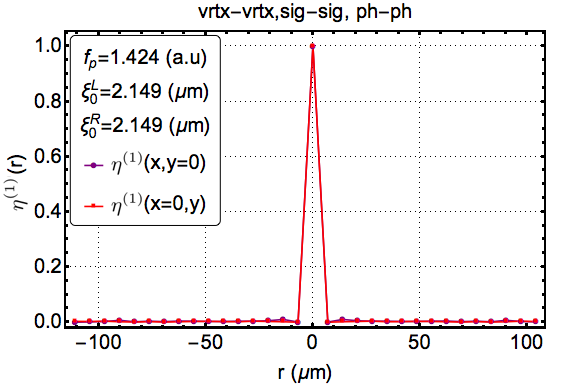}
			\includegraphics[width=0.35\textwidth]{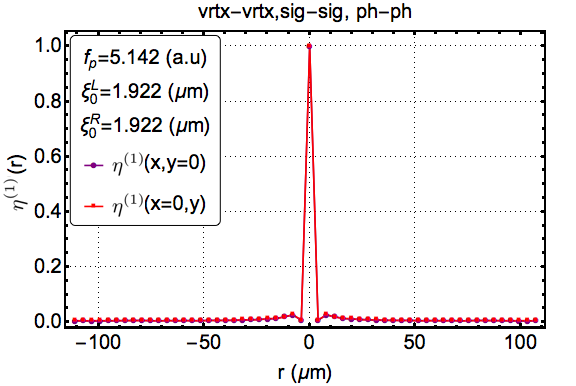}
			\includegraphics[width=0.35\textwidth]{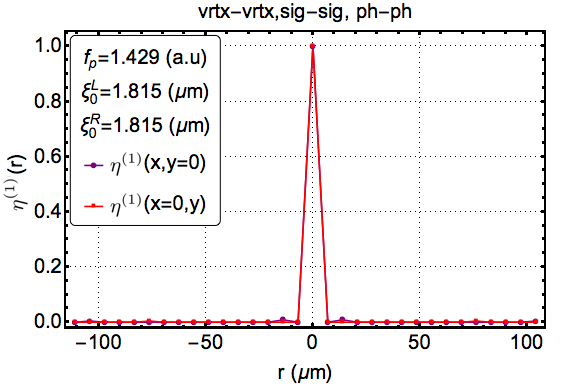}
			\includegraphics[width=0.35\textwidth]{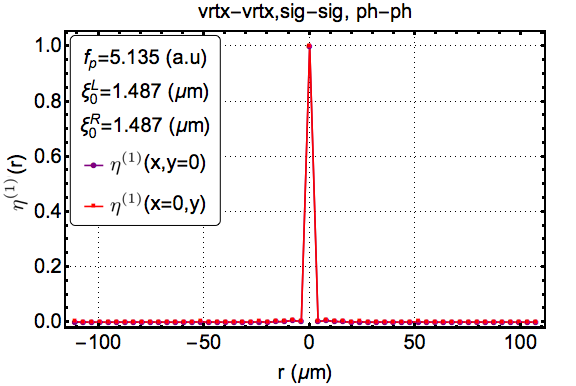}
		\end{center}
		\caption{Cross sections of vortex-vortex correlations in the signal mode of the photonic component. Varying pump power across the thresholds, the top two rows depict disordered states, the third row the thresholds, and bottom rows show the ordered states, for the lower (left column) and upper (right column) OPO thresholds.}
		\label{figSM:4}
	\end{figure*}
	
	\section{Vortex-vortex  correlations}
	As discussed in the main paper, (anti)vortex-(anti)vortex  and vortex-antivortex correlations defined as in Eq.~\red{3} of the main script, are helpful to the interpretation of the vortex gas distribution between different vortex charges, modes and components. These correlations are discussed in detail in the main paper.  
	In the following, we discuss a few additional aspects:
		
	\subsection{Vortex-vortex and antivortex-antivortex correlations in the same mode, component and threshold}

		Vortex-vortex and antivortex-antivortex correlations are reported in Figs.~\ref{figSM:2} and \ref{figSM:3}. As expected these functions are peaked at the origin indicating autocorrelation. There is then a minimum at very short distances, indicating vortex-vortex (antivortex-antivortex) repulsion.  
		In the disordered states, the healing lengths and number of vortices are large,
		therefore one finds a plateau in correlations at large distances which, assuming the presence of vortices, verifies the random nature of the vortex gas distribution with the presence of dipoles and the absence of vortex clustering.
		In the ordered states there are no correlations between vortices with the same charge in the same mode of the same component, suggesting an absence of vortices, apart from a finite and small correlations  in the photonic idler in both UT and LT (lower row), indicating the presence of a few vortices with the same sign even at large distances.
		In the excitonic component, correlations in the ordered phase are zero in all cases (except for the autocorrelation  at the origin).
		In Fig.~\ref{figSM:4} we plot the cross-section of the top panels (signal, photonic) of Fig.~\ref{figSM:2} as a function of different pump strengths, showing the decrease of the large-distance plateau of the vortex correlator as the pump strength is varied going from the disordered to ordered phase across the BKT transition.
		
\begin{figure*} 
		\begin{center}
			\includegraphics[width=0.42\textwidth]{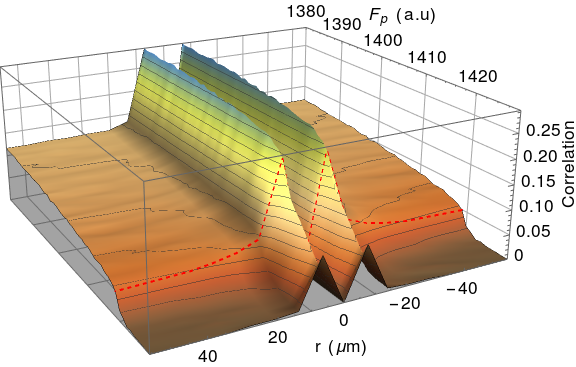}
			\includegraphics[width=0.42\textwidth]{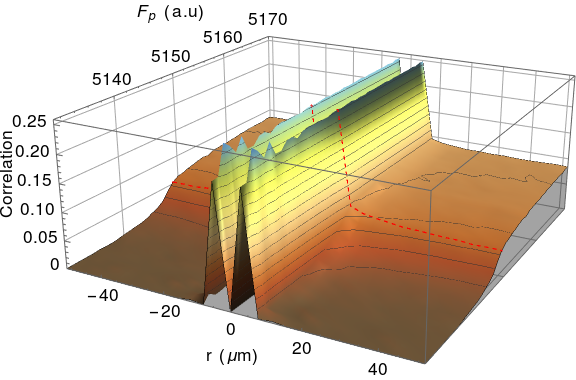}
			\includegraphics[width=0.42\textwidth]{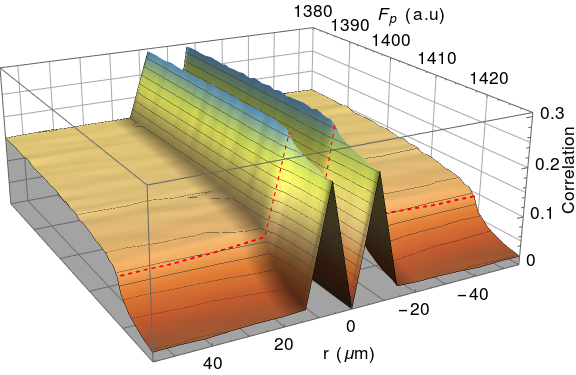}
			\includegraphics[width=0.42\textwidth]{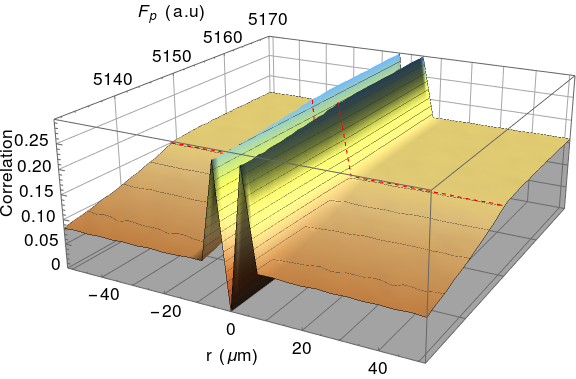}
			\includegraphics[width=0.42\textwidth]{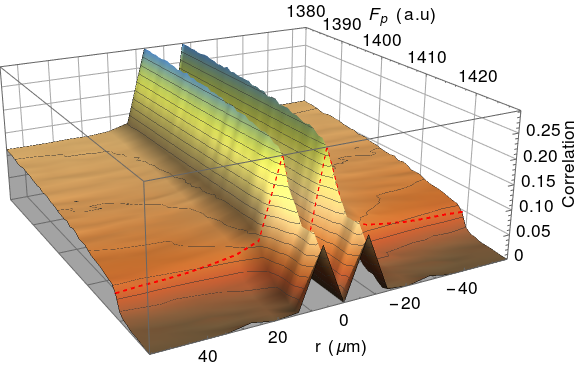}
			\includegraphics[width=0.42\textwidth]{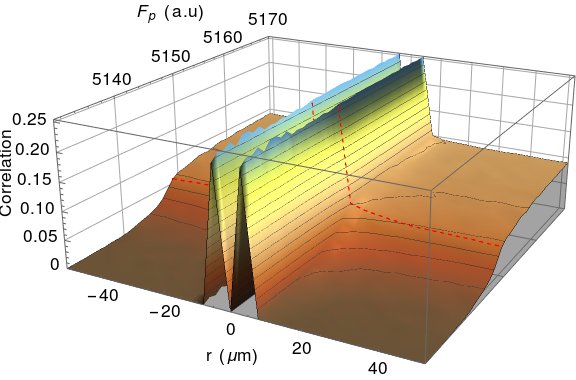}
			\includegraphics[width=0.42\textwidth]{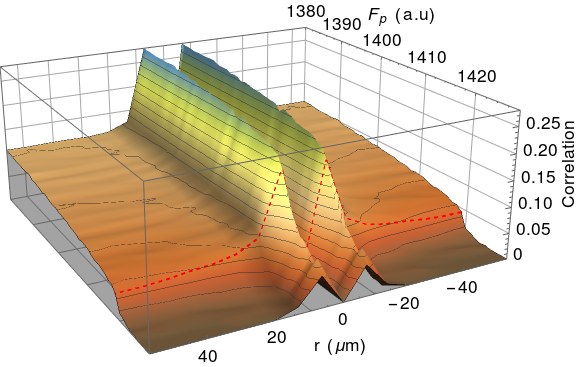}
			\includegraphics[width=0.42\textwidth]{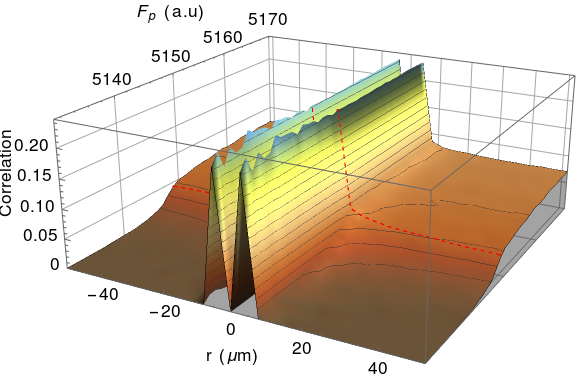}
		\end{center}
		\caption{Correlations between vortex-antivortex/antivortex-vortex in the same mode and component. From top to bottom:
			Vortex-antivortex correlations in the signal and idler modes of photons (top two rows) and excitons (bottom two rows)
			for lower (left column) and upper (right column) OPO thresholds.}
		\label{figSM:5}
	\end{figure*}

		Fig.~\ref{figSM:5} illustrates vortex-antivortex pairing in the different components; as discussed in the main paper, all components show a similar pairing crossover from a “randomly distributed” disordered phase to a “single-pair” ordered phase, except for the photonic idler (right column, second and fourth plot from the top).  

	Same- and opposite-sign correlators can be successfully employed to investigate the phase locking between OPO components; Fig.~\ref{figSM:7} and Fig.~\ref{figSM:8} show these correlators. As discussed in the main text, the inter-mode spatial pairing mechanism takes place only within modes belonging to the excitonic fraction, but not for the photonic ones.

	\subsection{Vortex-vortex correlation anisotropy}
	
\begin{figure*} 
		\begin{center}
			\includegraphics[width=0.35\textwidth]{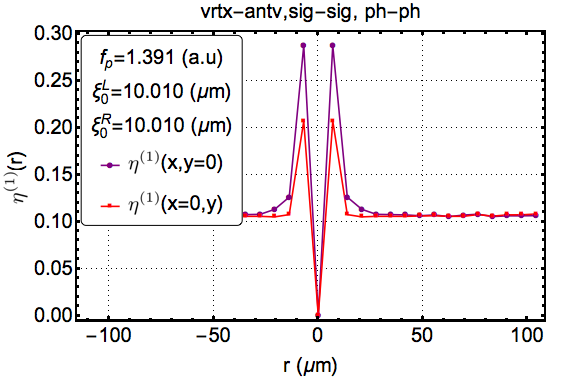}
			\includegraphics[width=0.35\textwidth]{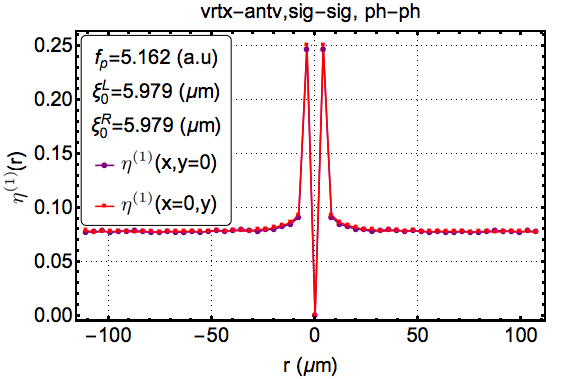}
			\includegraphics[width=0.35\textwidth]{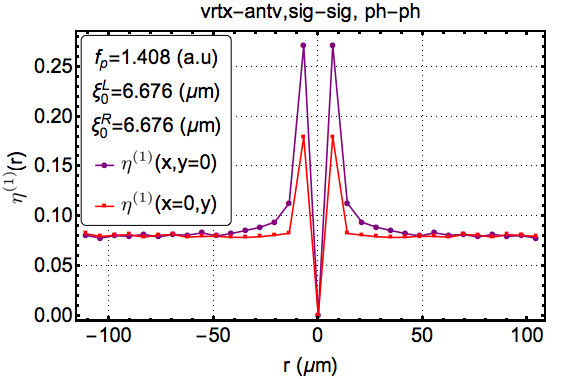}
			\includegraphics[width=0.35\textwidth]{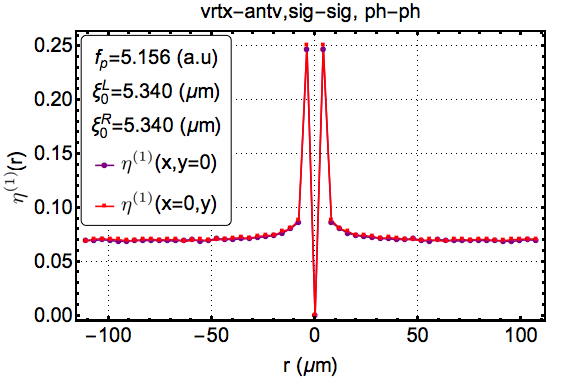}
			\includegraphics[width=0.35\textwidth]{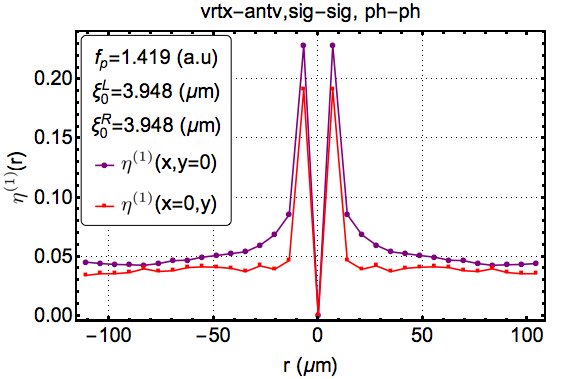}
			\includegraphics[width=0.35\textwidth]{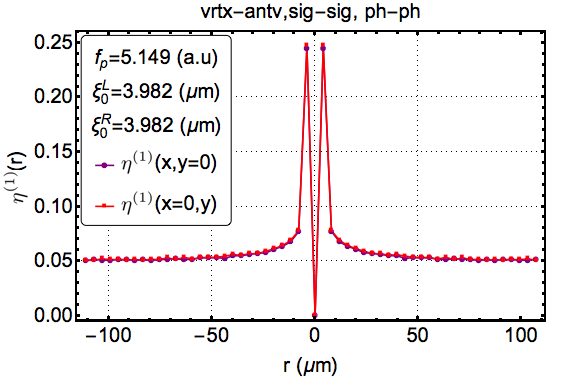}
			\includegraphics[width=0.35\textwidth]{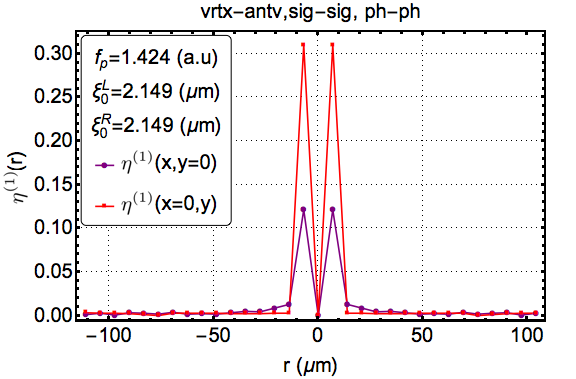}
			\includegraphics[width=0.35\textwidth]{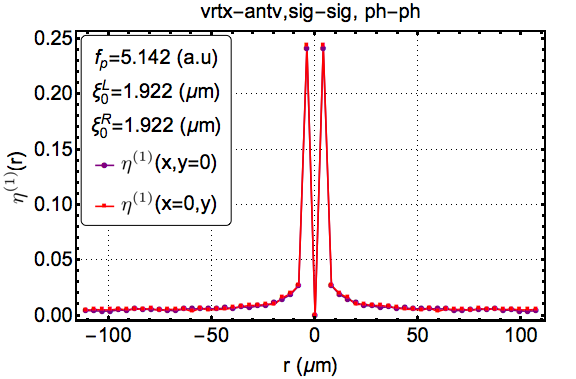}
			\includegraphics[width=0.35\textwidth]{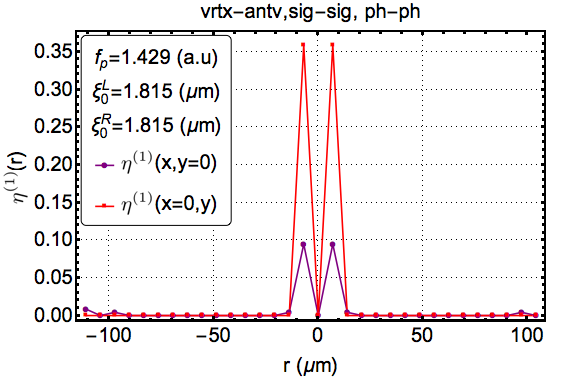}
			\includegraphics[width=0.35\textwidth]{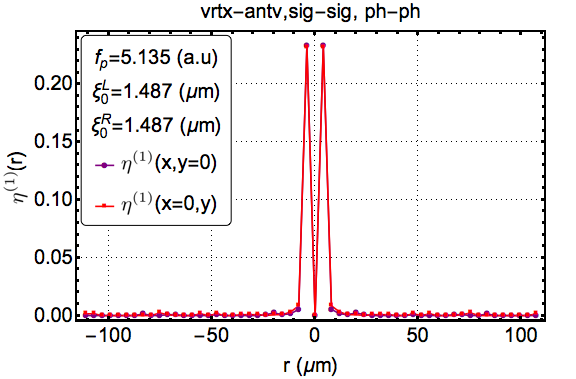}
		\end{center}
		\caption{
		Cross sections of vortex-antivortex correlations $\eta^{(1)}$ in the signal mode of the photonic component. Varying pump power across the thresholds, the top two rows depict disordered states, the third row the thresholds, and bottom rows show the ordered states, for the lower (left column) and upper (right column) OPO thresholds.}
		\label{figSM:6}
	\end{figure*}

\begin{figure*} 
		\begin{center}
			\includegraphics[width=0.42\textwidth]{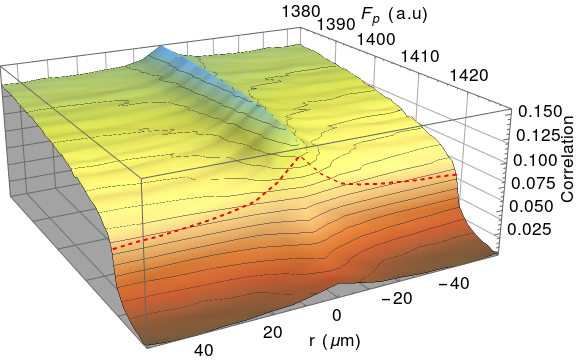}
			\includegraphics[width=0.42\textwidth]{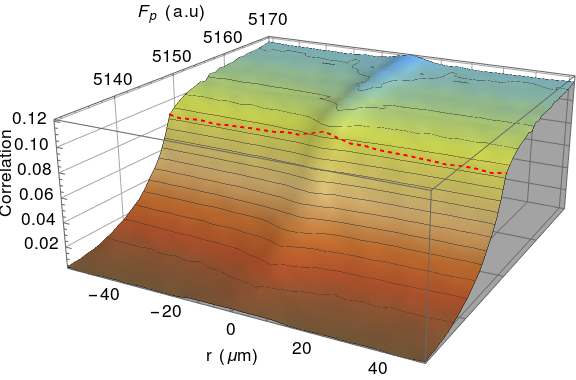}
			\includegraphics[width=0.42\textwidth]{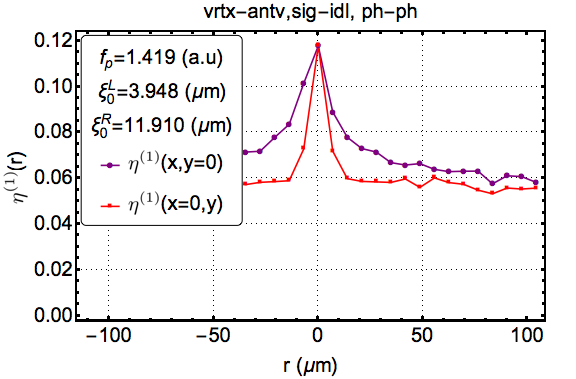}
			\includegraphics[width=0.42\textwidth]{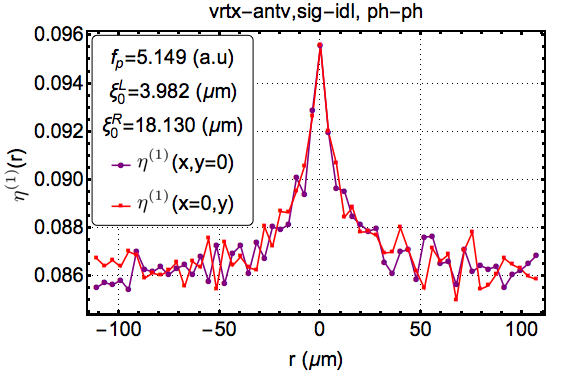}
			\includegraphics[width=0.42\textwidth]{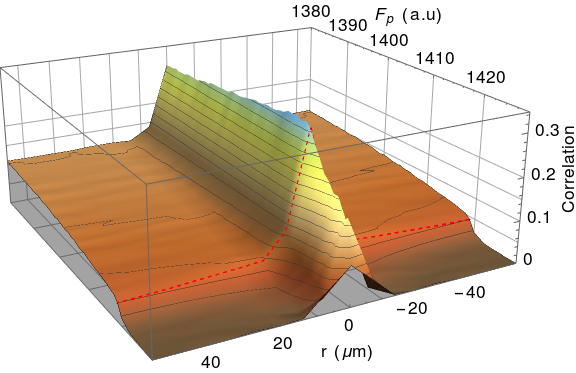}
			\includegraphics[width=0.42\textwidth]{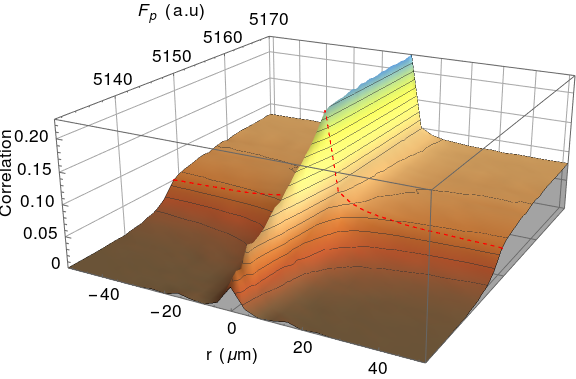}
			\includegraphics[width=0.42\textwidth]{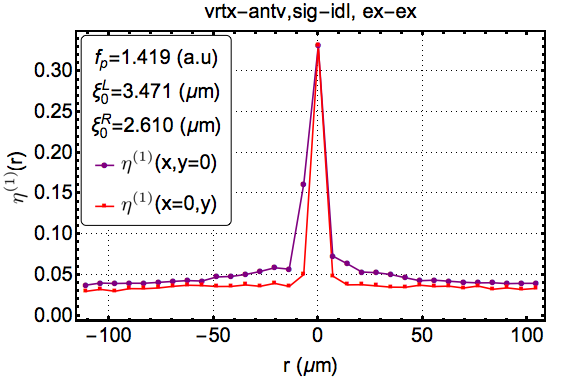}
			\includegraphics[width=0.42\textwidth]{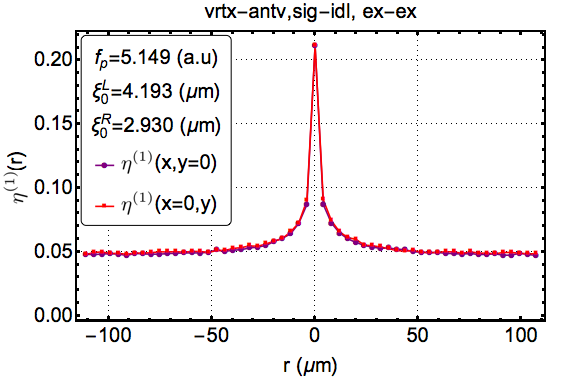}
		\end{center}
		\caption{Cross sections of vortex-antivortex correlations $\eta^{(1)}$  between signal and idler in the photonic (upper two rows) and excitonic (bottom two rows) components.}
		\label{figSM:7}
	\end{figure*}

\begin{figure*} 
		\begin{center}
			\includegraphics[width=0.42\textwidth]{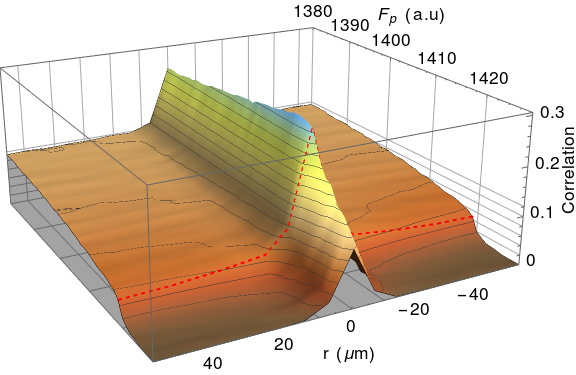}
			\includegraphics[width=0.42\textwidth]{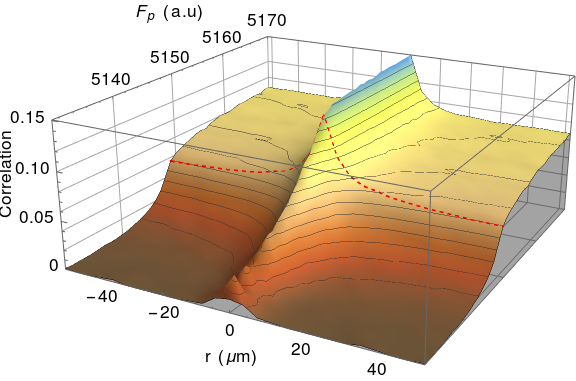}
			\includegraphics[width=0.42\textwidth]{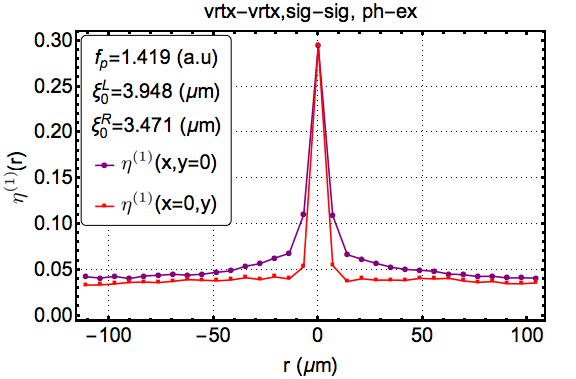}
			\includegraphics[width=0.42\textwidth]{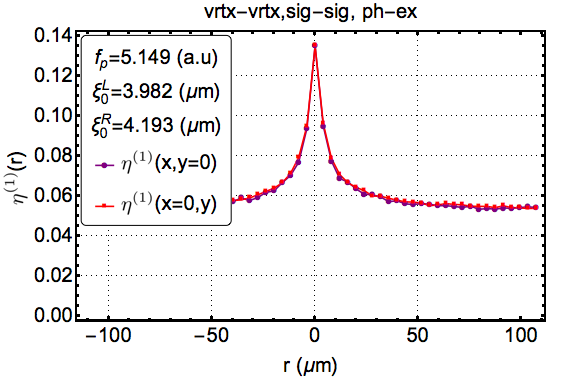}
			\includegraphics[width=0.42\textwidth]{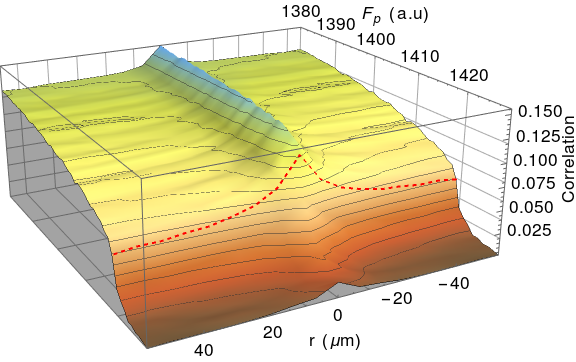}
			\includegraphics[width=0.42\textwidth]{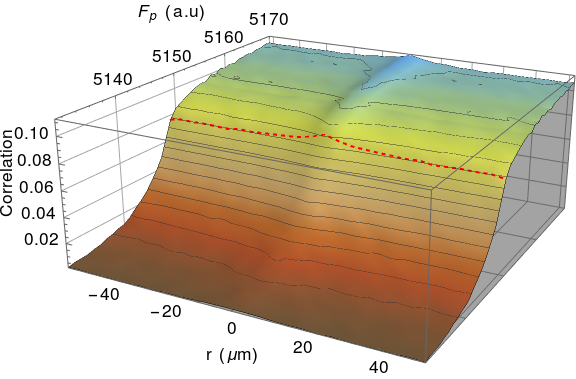}
			\includegraphics[width=0.42\textwidth]{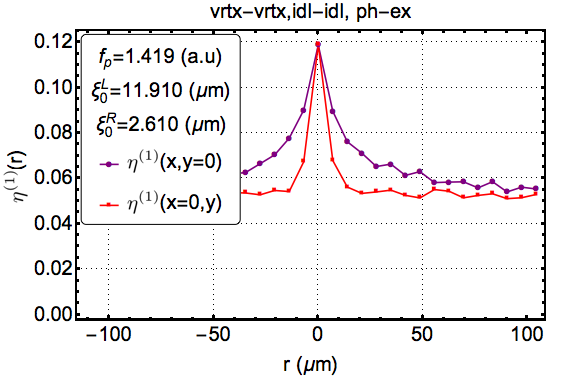}
			\includegraphics[width=0.42\textwidth]{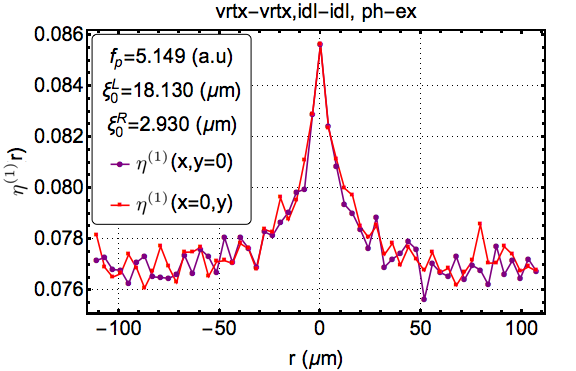}
		\end{center}
		\caption{Cross sections of vortex-vortex correlations $\eta^{(1)}$ between exciton and photon components in the signal (upper two rows) and idler (lower two rows) modes.}
		\label{figSM:8}
	\end{figure*}

	In Fig.~\ref{figSM:6} we plot the cross-section of vortex-antivortex correlations in the signal mode of the photonic component, as the pump is varied across the thresholds. Comparison between the LT and UP, left and right columns respectively, clearly shows an anisotropic behaviour which takes place only at the LT. 
	In order to explain this phenomenon, we consider possible currents in the frame of the signal mode originating from the OPO nature of the system. We rule out such an explanation by finding an average zero net current at all pump values across the phase transition.
	We further consider a possible anisotropic behaviour originating from the asymmetry of the diffusive and/or nonlinear coefficients of the 
	driven-dissipative equations for the phase~\cite{altman2015twodimensional} field, i.e. the Kardar–Parisi–Zhang equation~\cite{KPZ1986}. 
	We proceed by deriving such coefficients calculated from the extracted numerical $\textbf{k}_s$ from the mapping to the complex Ginzburg-Landau model~\cite{zamora2016driving}.
	The coefficients are reported in Tab.~\ref{TabCoeffKPZ}. We find that our parameters correspond to a weak anisotropy. It maybe possible that the anisotropy in the equation for the phase will result in the anisotropic interactions between vortices leading to the anisotropic correlations observed. It is, however, difficult to conclude whether this could justifies our hypothesis. As mentioned in the main text, the analytical estimates of forces between vortices exist for the isotropic case only \cite{wachtel2016electrodynamic}. Given our numerical fundings it would be of great interest to extend the analytical analysis of Ref \cite{wachtel2016electrodynamic} to the anisotropic case. 
	
	\begin{table}[ht]
		\centering

		\begin{tabular}[t]{lccccc}
			\hline
			$F_p$	& scaled $f_p$ &	$D_x$ &	$D_y$ &	$\lambda_x$ &	$\lambda_y$\\
			\hline
			1.391	& 0.01851	&0.0505&	0.0525&	-0.3282	&-0.3710\\
			1.408	&0.01873	&0.0515&	0.0548	&-0.3276	&-0.3690\\
			1.433	&0.01906	&0.0530&	0.0583	&-0.3267	&-0.3659 \\
			\hline
		\end{tabular}
				\caption{Extracted KPZ coefficients for $\textbf{k}_{\mathrm{C},s} = 0.6136 \mathrm{\mu m^{-1}}$}
				\label{TabCoeffKPZ}
	\end{table}%

\subsection{Real-space phase profiles and first-order correlation function $g^{(1)}$}
	
\begin{figure*} 
		\begin{center}
			\includegraphics[width=0.42\textwidth]{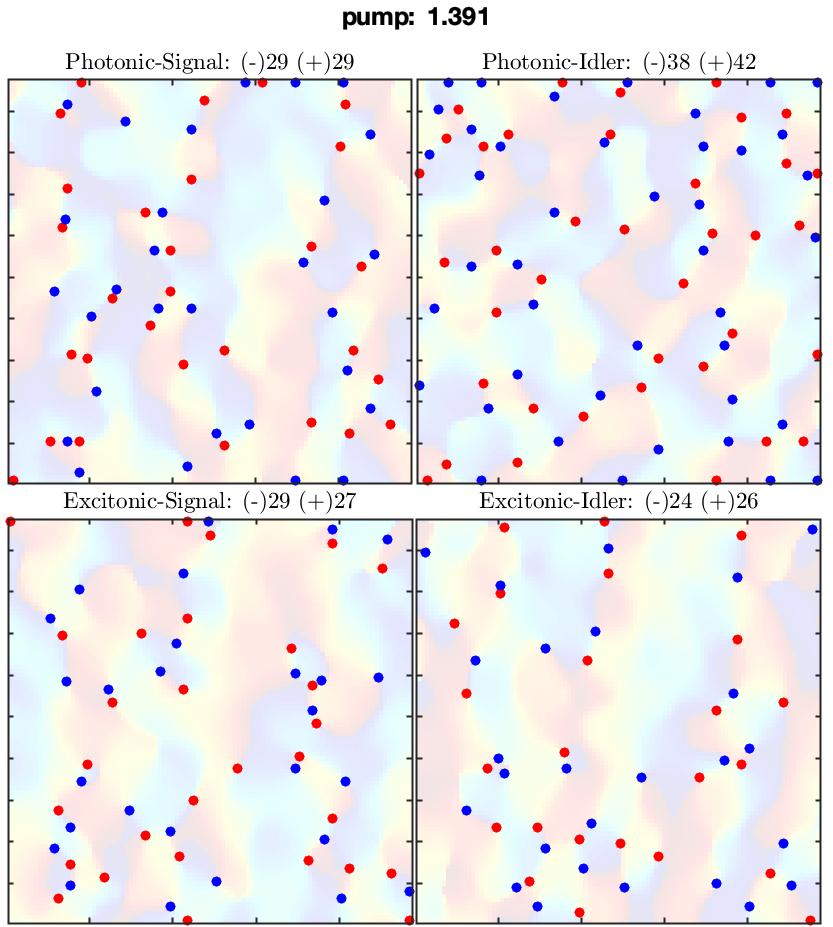}
			\includegraphics[width=0.42\textwidth]{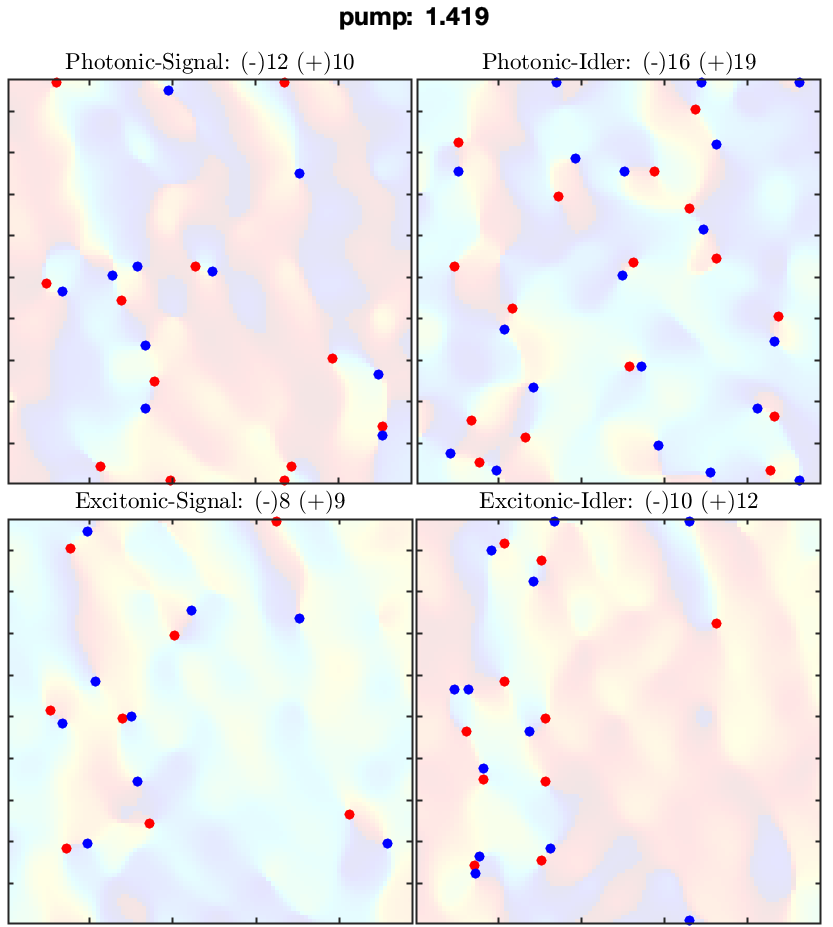}
			\\
			\vspace{4mm}
			\includegraphics[width=0.42\textwidth]{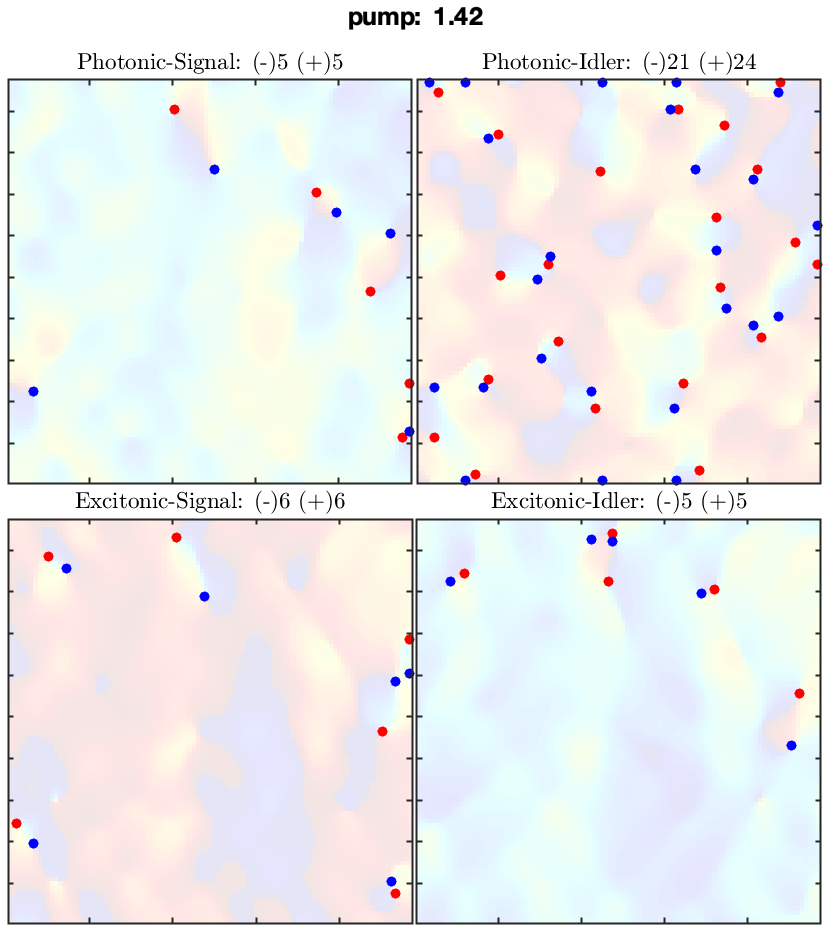}
			\includegraphics[width=0.42\textwidth]{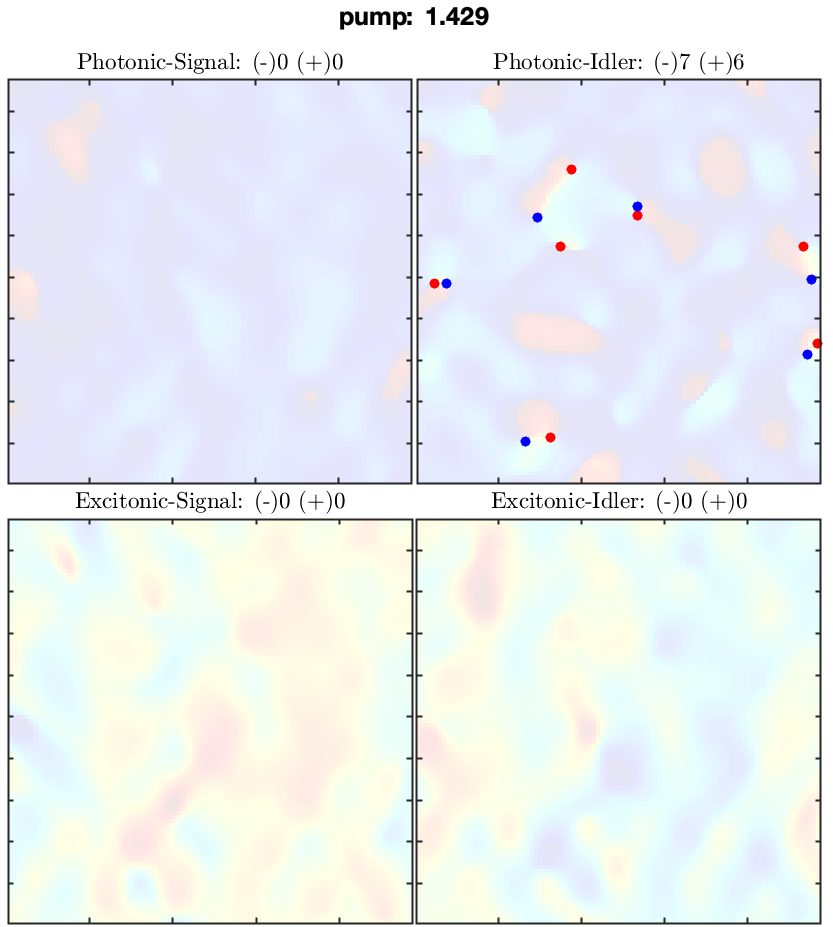}
		\end{center}
		\caption{Spatial phase profiles of all four components across the lower threshold. Only a zoomed-in portion with area $L/8 \times L/8$ is shown, where $L \times L$ is the total size of the sample. Top: In disordered states, far from (left)
			and near  (right) the transition. Bottom: In ordered states, near to (left) and far from (right) the transition.}
		\label{figSM:snapsLT}
	\end{figure*}

\begin{figure*} 
		\begin{center}
			\includegraphics[width=0.42\textwidth]{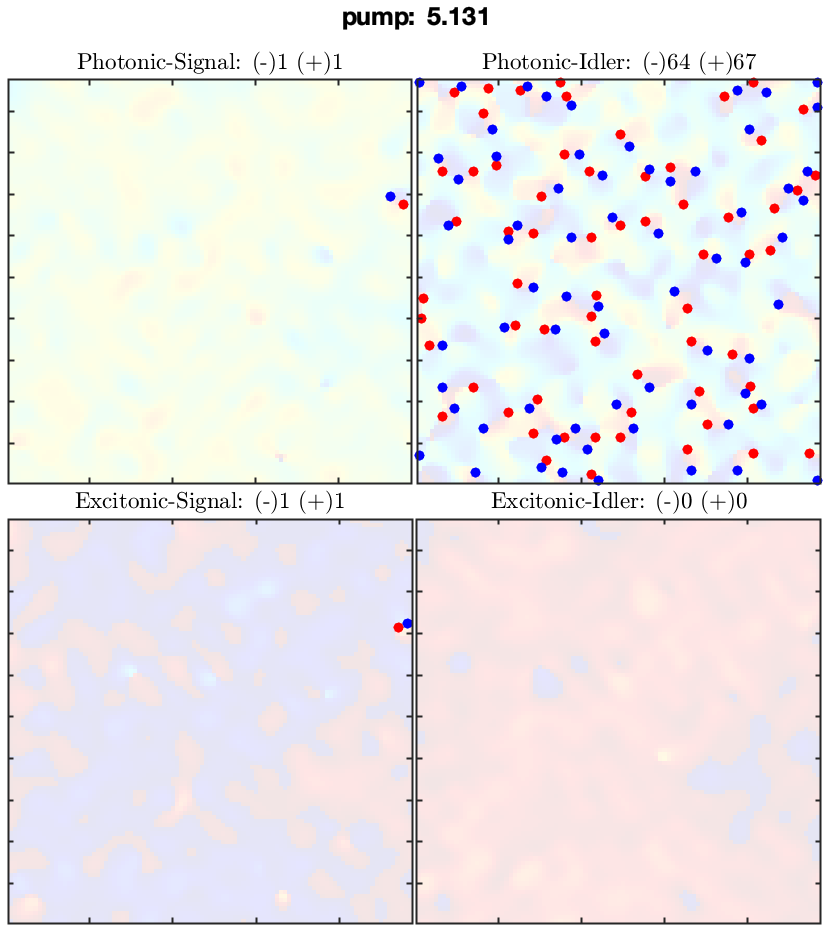}
			\includegraphics[width=0.42\textwidth]{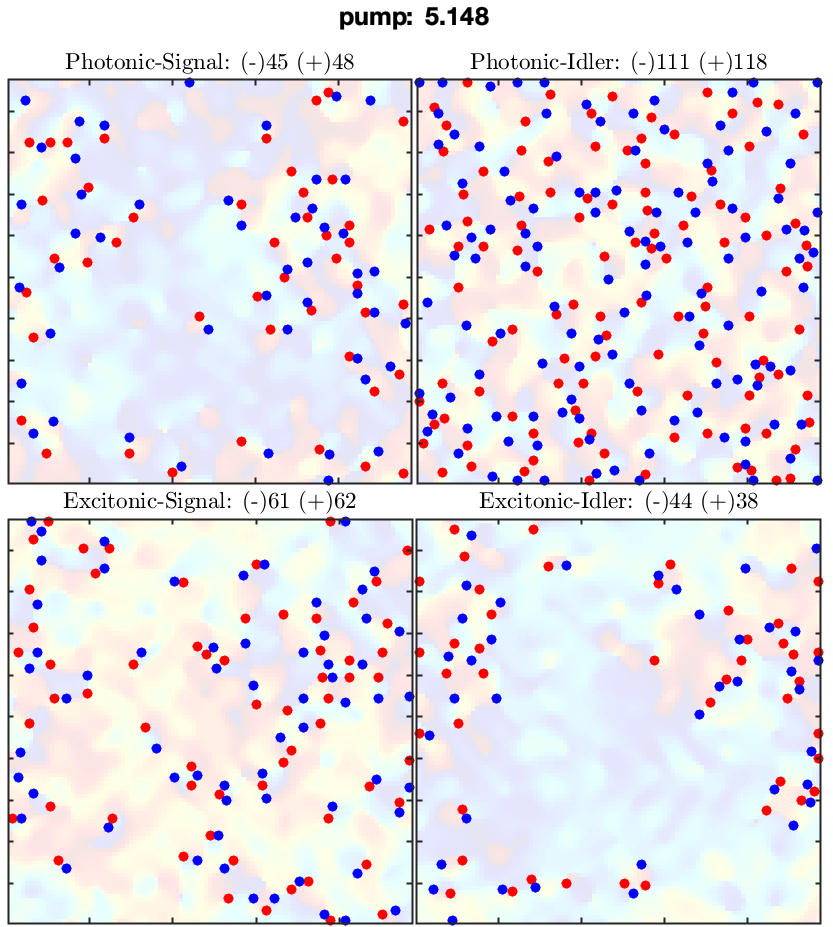}
			\\
			\vspace{4mm}
			\includegraphics[width=0.42\textwidth]{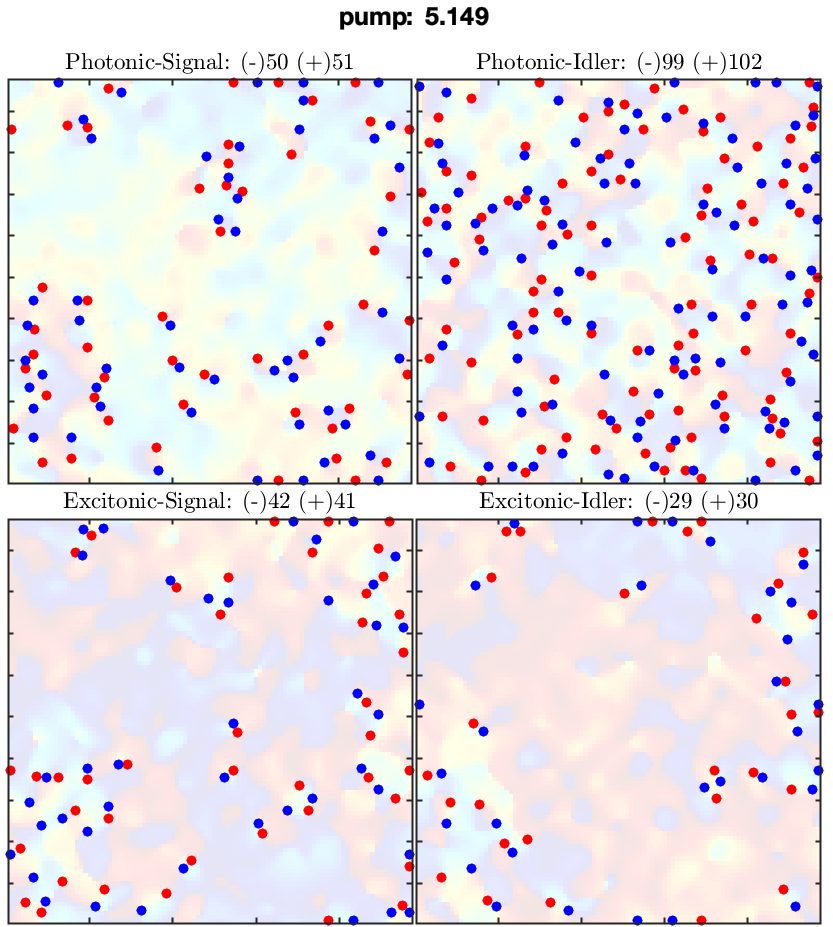}
			\includegraphics[width=0.42\textwidth]{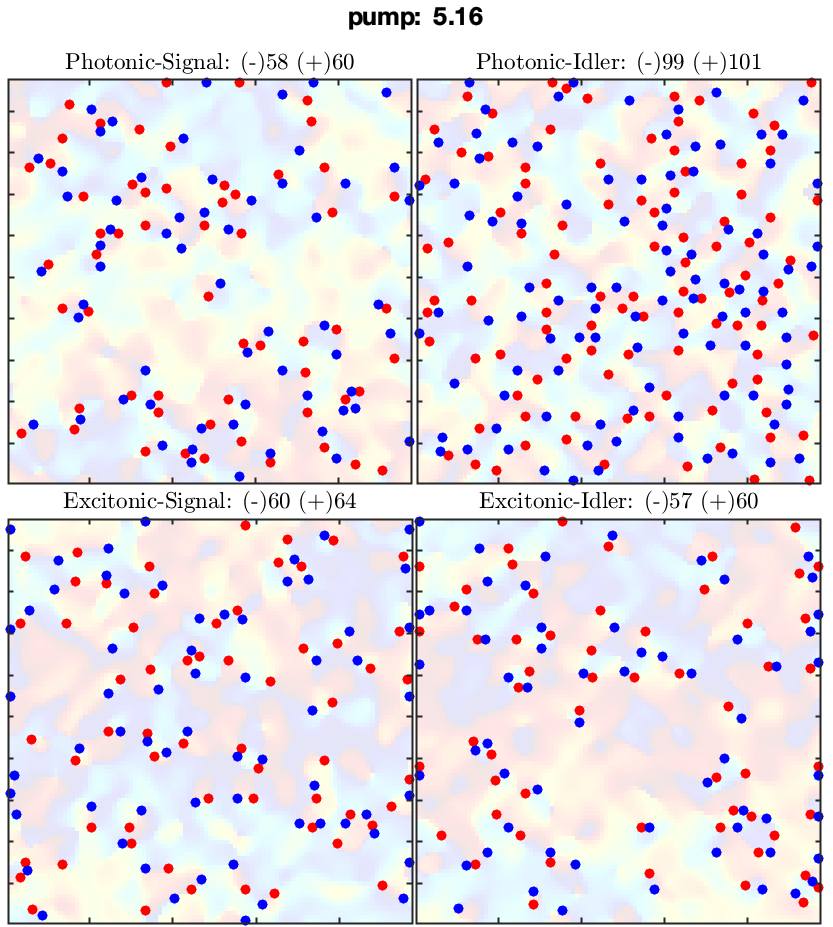}
		\end{center}
		\caption{As in Fig.~\ref{figSM:snapsLT}, but for the upper threshold}
		\label{figSM:snapsUT}
	\end{figure*}

	In Figs.~\ref{figSM:snapsLT} and \ref{figSM:snapsUT} we show the phase profiles of a single realisation of the polariton system for different pumps at the LT and UT respectively. 
	On top of each plot, we draw a blue (red) dot to locate a topological defect with positive (negative) circulation sign.
	Numerical methods for the detection and counting of vortices are reported in detail in Refs.~\cite{comaron2018dynamical,Comaron2021}.
	The total number of vortices with positive ($+$) and negative ($-$) circulation is indicated at the top of each panel.
	
	In both the LT and UT crossovers, the pairing process is clearly visible: it increases as the system moves towards the ordered phase. 
	Comparison between different components shows different configurations. 
	Interestingly, we note that deep in the quasi-ordered phase, it is only in the photonic-idler mode that V-AV pairs survive in a significant number. 
	For the latter case, the system exhibits the presence of ensembles  of topological defects.
	The vortex distribution is found to be mainly composed by V-AV pairs. Scanning across many single realisations, we also find trace evidence of free vortices, which are however appearing rarely when one takes into account all realisations.
	We believe that this rate is insufficient to alter the power-law behaviour of the first-order correlator discussed in the main manuscript and below, which is an average over all realisations.
	
\begin{figure*} 
		\begin{center}
			\includegraphics[width=0.42\textwidth]{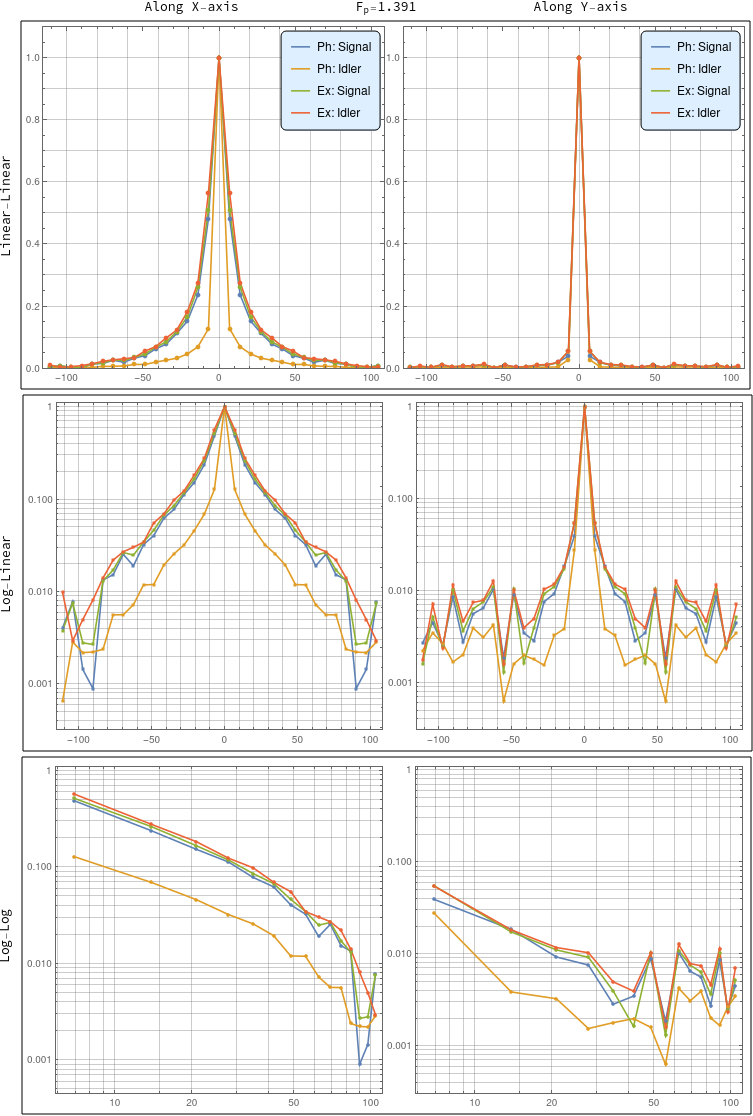}
			\includegraphics[width=0.42\textwidth]{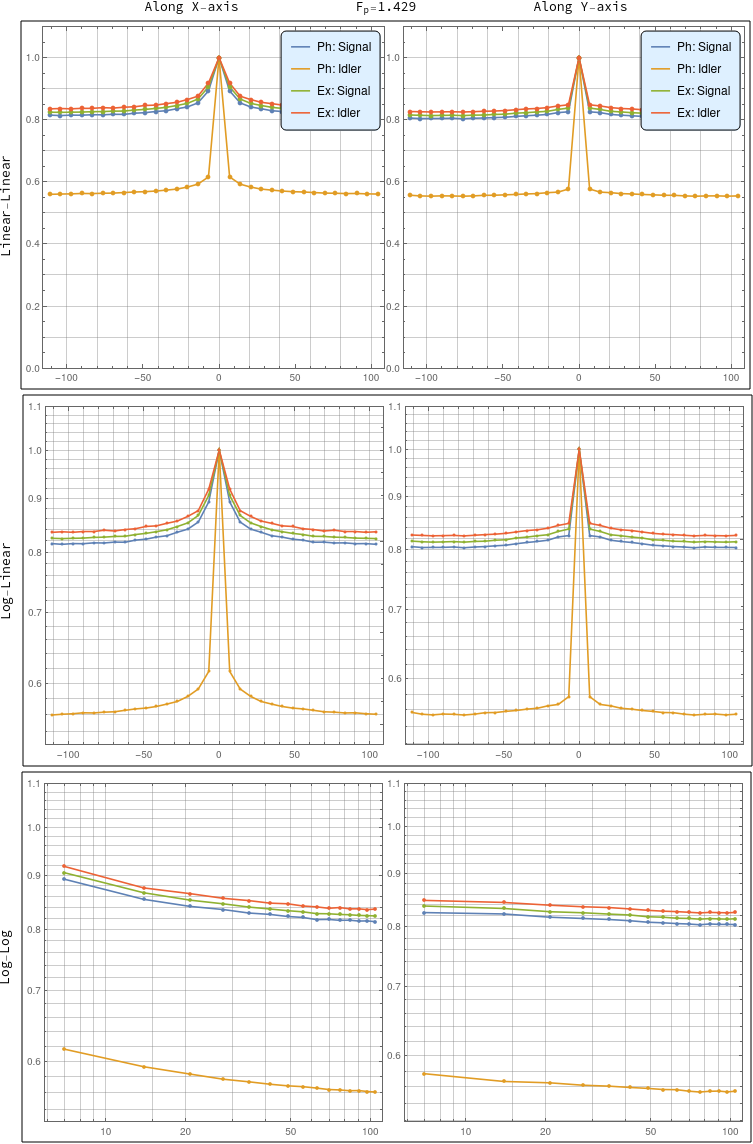}
			\\
			\vspace{4mm}
			\includegraphics[width=0.42\textwidth]{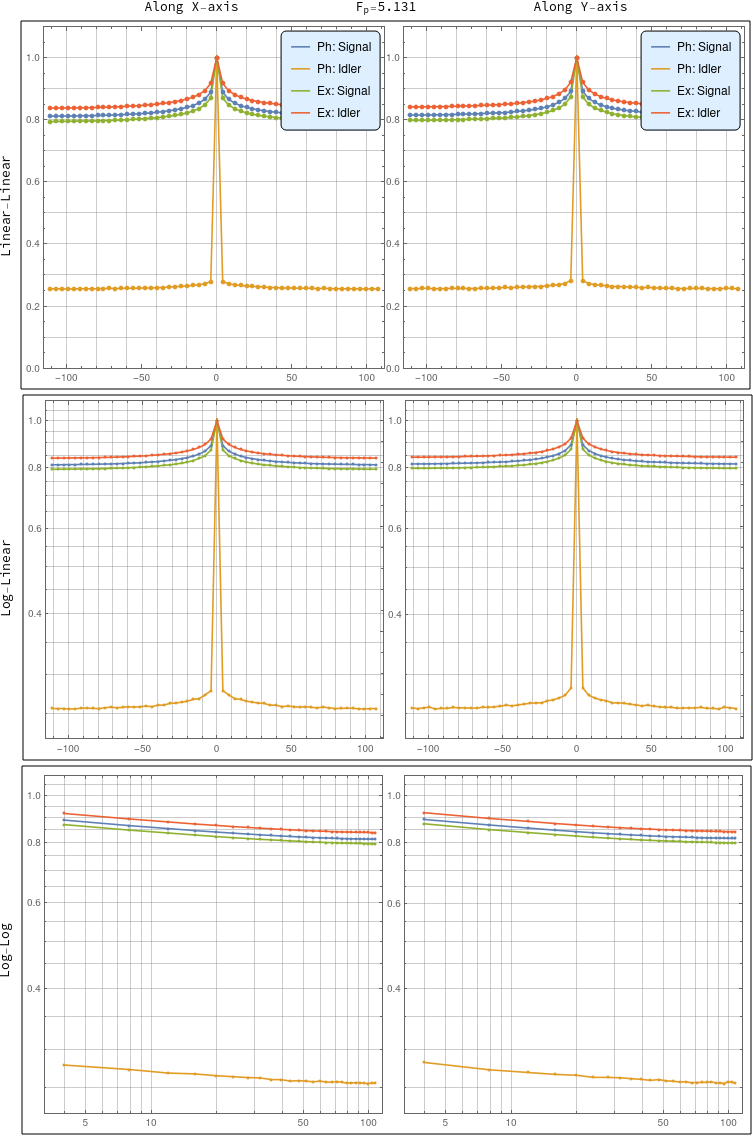}
			\includegraphics[width=0.42\textwidth]{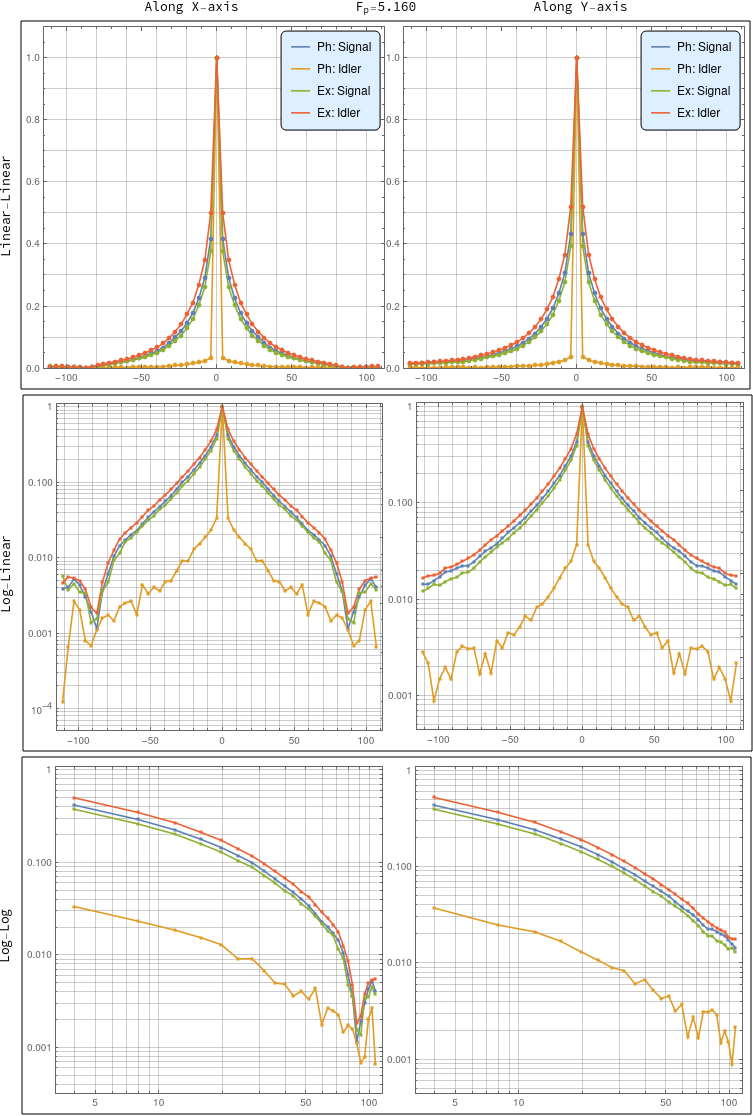}
		\end{center}
		\caption{First-order correlation function, i.e. Eq.~\red{(3)} of the main text, shown in different scales, around the LT (upper row) and UT (bottom row) for different pump powers deep in the disordered and ordered phases.}
		\label{figSM:g1}
	\end{figure*}

	We proceed investigating the behaviour of phase fluctuations by computing spatial correlations, i.e.~Eq.~\red{(3)} of the main paper.
	For pump powers well below and above the critical LT and UT, in Fig.~\ref{figSM:g1} we plot the first-order correlation function along the $x$ and $y$ axes for all components. 
	Note the different plots in linear-linear, linear-logarithmic, and logarithmic-logarithmic scales. 
	Let us focus on the ordered-phase cases. At very short distances, spatial correlations decay at different rates and eventually plateau at different values: from this one can infer that short range (high momentum) fluctuations are dependent on microscopics and vary between components.
	At large distances, instead, the plots clearly exhibit similar decay of correlations in the different components; this is visible both in the linear and logarithmic scaled plots. 
	Quantifying this behaviour is done by the extracting the power-law exponents as shown in Fig.~\red{3} and discussed in the text of the main paper.
	Summarising, these different features suggest that, in the ordered phase, presence  of vortices in the photonic-idler component destroy coherence at small distances, 
	which is however recovered at long distances given that the different components are strongly phase-locked to one another.

